\documentclass[12pt,a4paper]{article}
\usepackage[T1]{fontenc}
\usepackage[sc,osf]{mathpazo}
\usepackage{a4wide}  
\usepackage{latexsym,amsthm,amsfonts,amsmath,mathrsfs,amssymb}
\usepackage{booktabs} 
\usepackage{ifpdf}
\ifpdf
\usepackage[pdftex,unicode,implicit]{hyperref}
\hypersetup{%
  pdftitle    = {Non-Abelian black string solutions of N=(2,0),d=6 supergravity}
  pdfkeywords = {Yang-Mills, black string, supergravity, supersymmetry,
    dimensional reduction, oxidation},
  pdfauthor   = {Pablo A. Cano, Tomas Ortin, Camilla Santoli},
  plainpages  = true,
  colorlinks  = true,
  citecolor   = blue,
  urlcolor    = red,
  linkcolor   = black
}
\newcommand{\hepth}[1]{{\tt
\href{http://www.arXiv.org/abs/hep-th/#1}{hep-th/#1}}}

\newcommand{\arxiv}[1]{{\tt arXiv:\href{http://www.arXiv.org/abs/#1}{#1}}}
\newcommand{\doi}[1]{{\tt DOI:\href{http://dx.doi.org/#1}{#1}}}
\else
  \usepackage[dvips]{graphicx}
  \usepackage[unicode,implicit]{hyperref}
  \newcommand{\hepth}[1]{{\tt hep-th/#1}}

  \newcommand{\arxiv}[1]{{\tt arXiv:#1}}
  \newcommand{\doi}[1]{{\tt DOI:#1}}
\fi
\makeatletter
\@addtoreset{equation}{section}
\makeatother

\begin{document}

\begin{flushright}
\small
IFT-UAM/CSIC-16-058\\
\texttt{arXiv:1607.02595 [hep-th]}\\
July 9\textsuperscript{th}, 2016\\
\normalsize
\end{flushright}

\vspace{1cm}

\begin{center}

{\Large {\bf {Non-Abelian black string solutions\\[.5cm] of $\mathcal{N}=(2,0),d=6$ supergravity}}}
 
\vspace{1.5cm}

\renewcommand{\thefootnote}{\alph{footnote}}
{\sl\large  Pablo A.~Cano}${}^{1,}$\footnote{E-mail: {\tt pabloa.cano [at] estudiante.uam.es}}
{\sl\large  Tom\'{a}s Ort\'{\i}n}${}^{1,}$\footnote{E-mail: {\tt Tomas.Ortin [at] csic.es}}
{\sl\large and Camilla Santoli}${}^{2,}$\footnote{E-mail: {\tt Camilla.Santoli [at] mi.infn.it}}

\setcounter{footnote}{0}
\renewcommand{\thefootnote}{\arabic{footnote}}

\vspace{1cm}

${}^{1}${\it Instituto de F\'{\i}sica Te\'orica UAM/CSIC\\
C/ Nicol\'as Cabrera, 13--15,  C.U.~Cantoblanco, E-28049 Madrid, Spain}\\ \vspace{0.3cm}

${}^{2}${\it Dipartimento di Fisica, Universit\`a di Milano, and
INFN, Sezione di Milano, Via Celoria 16, I-20133 Milano, Italy.
}

\vspace{1cm}


{\bf Abstract}

\end{center}

\begin{quotation}
  {\small We show that, when compactified on a circle, $\mathcal{N}=(2,0),d=6$
    supergravity coupled to 1 tensor multiplet and $n_{V}$ vector multiplets
    is dual to $\mathcal{N}=(2,0),d=6$ supergravity coupled to just
    $n_{T}=n_{V}+1$ tensor multiplets and no vector multiplets. Both theories
    reduce to the same models of $\mathcal{N}=2,d=5$ supergravity coupled to
    $n_{V5}=n_{V}+2$ vector fields. We derive Buscher rules that relate
    solutions of these theories (and of the theory that one obtains by
    dualizing the 3-form field strength) admitting an isometry. Since the
    relations between the fields of $\mathcal{N}=2,d=5$ supergravity and those
    of the 6-dimensional theories are the same with or without gaugings, we
    construct supersymmetric non-Abelian solutions of the 6-dimensional gauged
    theories by uplifting the recently found 5-dimensional supersymmetric
    non-Abelian black-hole solutions. The solutions describe the usual
    superpositions of strings and waves supplemented by a BPST instanton in
    the transverse directions, similar to the gauge dyonic string of Duff,
    L\"u and Pope. One of the solutions obtained interpolates smoothly between
    two AdS$_{3}\times$ S$^{3}$ geometries with different radii.}

\end{quotation}

\newpage
\pagestyle{plain}

\tableofcontents


\section*{Introduction}

The supergravity theories with 8 real supercharges provide a very interesting
arena for the construction and study of supersymmetric solutions because they
have enough symmetry to be tractable and exhibit interesting properties such
as the attractor mechanism of their black-hole and black-string solutions
\cite{Ferrara:1995ih,Strominger:1996kf,Ferrara:1996dd,Ferrara:1996um,Ferrara:1997tw}
but not so much symmetry that only a few models are permitted.\footnote{A
  general but deep review of all these theories can be found in
  Ref.~\cite{Freedman:2012zz} and for the 4-dimensional case, only, in
  Ref.~\cite{Andrianopoli:1996cm}. The 4- and 5-dimensional ones are also
  reviewed in Ref.~\cite{Ortin:2015hya}, with emphasis on the supersymmetric
  bosonic solutions.}  

Most of the work on these theories has been devoted to the 4-and 5-dimensional
ones for different reasons: for a given matter content many models possible;
they are the effective theories of type~II superstrings compactified on
Calabi-Yau 3-folds (times a circle in the 4-dimensional case); they have rich
geometrical structures known as Special Geometry (K\"ahler in $d=4$, real in
$d=5$); they admit supersymmetric black-hole solutions etc.  In fact, most of
whose supersymmetric solutions have been classified in
Refs.~\cite{Tod:1983pm,Caldarelli:2003pb,Meessen:2006tu,Huebscher:2006mr,Cacciatori:2008ek,Hubscher:2008yz,Meessen:2012sr}
and
Refs.~\cite{Gauntlett:2002nw,Gauntlett:2003fk,Gauntlett:2004qy,Gutowski:2004yv,Gutowski:2005id,Bellorin:2006yr,Bellorin:2007yp,Bellorin:2008we}
respectively.  

Much less work has been done in the 6-dimensional theories (often called
$\mathcal{N}=(2,0),d=6$ supergravities because they have chiral fermions),
whose structure is not as rich and which are not associated to Calabi-Yau
compactifications. The pure supergravity theory, first constructed in
Ref.~\cite{Marcus:1982yu} by dimensional reduction from 11-dimensional
supergravity \cite{Cremmer:1978km} contains the graviton, gravitino and a
2-form with anti-selfdual 3-form field strength and it does not admit a
covariant action, which makes it more complicated to work with. This theory
can be coupled to vector multiplets (which have no scalars), tensor multiplets
(which have real scalars which always parametrize the same symmetric space
$\mathrm{SO}(1,n_{T})/\mathrm{SO}(n_{T})$ and 2-forms whose 3-form field
strengths are selfdual) and hypermultiplets (with scalars that parametrize
arbitrary quaternionic-K\"ahler manifolds). One way to avoid the complications
of having to deal with chiral 2-forms\footnote{That is: 2-form potentials with
  selfdual or anti-selfdual 3-form field strengths.} is to consider theories
with just one tensor multiplet so the two chiral 2-forms of opposite
chiralities combine into one unconstrained 2-form. These theories can describe
the effective theory of the truncated, toroidally compactified Heterotic
String (metric, Kalb-Ramond 2-form and dilaton) and, coupled to vector
multiplets and hypermultiplets were constructed in
Refs.~\cite{Nishino:1984gk,Bergshoeff:1985mz,Nishino:1986dc}. The coupling to
an arbitrary number of tensor multiplets was described in
Ref.~\cite{Romans:1986er} and has attracted much less attention because it has
not been identified as the effective field theory of some string or M-theory
compactification yet and it cannot be gauged, at least in any conventional
sense, because it does not have vectors that can be used as gauge fields. The
coupling to tensors, vectors and hypermultiplets with some gaugings was
described in Ref.~\cite{Nishino:1997ff}, which is the reference that we are
going to use here.

The supersymmetric solutions of most of these theories have not yet been
classified either. The only $\mathcal{N}=(2,0),d=6$ supergravity theories
considered have been the pure supergravity theory in
Refs.~\cite{Gutowski:2003rg,Chamseddine:2003yy} and a theory with one tensor
multiplet and a triplet of vector multiplets with $\mathrm{SU}(2)$ and
$\mathrm{U}(1)$ gaugings via Fayet-Iliopoulos terms in
Ref.~\cite{Cariglia:2004kk}.

In this paper we are going to study the often disregarded
$\mathcal{N}=(2,0),d=6$ supergravity theories that have several tensor
multiplets with or without vector multiplets as a preparation to classify
their supersymmetric solutions and to study how these solutions are related to
the supersymmetric solutions of the $\mathcal{N}=2,d=5$ theories by
dimensional reduction on a circle \cite{kn:CMOT}. We are also going to use
these results to construct new supersymmetric solutions of the
$\mathcal{N}=(2,0),d=6$ supergravity theories in absence of a classification.

Let is explain how we intend to achieve these goals.

In general, the supersymmetric solutions of theories related by dimensional
reduction are also related: all the supersymmetric solutions of the
lower-dimensional theory can be uplifted to supersymmetric solutions of the
higher-dimensional theory while all the supersymmetric solutions of the
higher-dimensional theory admitting translational isometries
\cite{Bergshoeff:1994cb}\footnote{In the case of toroidal
  compactification. The general condition is that the Killing spinors of the
  higher-dimensional solutions can also be understood as spinors of the
  lower-dimensional theory. This requires the spinors to have a particular
  dependence (or independence) on the coordinates of the compactification
  manifold which, in turn, requires the solution to meet certain
  conditions. In toroidal compactifications the isometries associated to the
  circles must act without fixed points (be translational isometries). In more
  general cases the conditions have not been studied. Observe that this
  possible problem only arised in the dimensional reduction and never in the
  oxidation because, by assuming the lower-dimensional solution to be
  supersymmetric we are assuming the problem has not arisen in the reduction
  and the lower-dimensional solution has been obtained froma supersymmetric
  higehr-dimensional solution. From a more general perspective: dimensional
  reduction can break symmetries but dimensional oxidation can never do that.}
can also be reduced along the associated directions to supersymmetric
solutions of the lower-dimensional theories. Thus, one can get new
supersymmetric solutions of one of the theories from known supersymmetric
solutions of the other one.\footnote{Of course, the same can be done with
  non-supersymmetric solutions.} The basic reason for this correspondence is
that the Killing spinor equations of the higher-dimensional theory always give
the Killing spinor equations of the lower-dimensional one and, if the latter
admit solutions, also do the former. As explained in the footnote, it may not
be true the other way around.

Two conditions have to be met in order to apply this simple
solution-generating technique:

\begin{enumerate}
\item One has to know which theories are related by dimensional reduction.
\item The detailed relation (``dictionary'') between the fields of the higher- and
  lower-dimensional theories must also be known.
\end{enumerate}

In our case it does not seem to be widely known which models of
$\mathcal{N}=2,d=5$ supergravity are related by dimensional reduction to which
models of $\mathcal{N}=(2,0),d=6$ supergravity theories, actually. Thus, our
first task (Section~\ref{reduction}) will be to perform the dimensional
reduction of a general, ungauged, $\mathcal{N}=(2,0),d=6$ supergravity theory
with an arbitrary number of tensor and vector multiplets\footnote{The
  hypermultiplets do not couple to the vector and tensor multiplets and,
  clearly, their reduction leads to 5-dimensional hypermultiplets with exactly
  the same quaternionic-K\"ahler geometry.} to $d=5$ and identify to which
model of $\mathcal{N}=2,d=5$ supergravity. A careful identification of the
5-dimensional fields will provide us with the dictionary we need to reduce and
uplift solutions (Section~\ref{sec-uplifting}).

The identification of the 5-dimensional models leads to a surprise: there are
two different families of models of $\mathcal{N}=(2,0),d=6$ supergravity
related to the same family of models of $\mathcal{N}=2,d=5$ supergravity: the
family of models with 1 tensor multiplet and $n_{V}$ vector multiplets (that
we are going to call $\mathcal{N}=2A,d=6$ theories)\footnote{These are the
  theories related to the toroidal compactification and truncation of the
  Heterotic String. We also consider the 6-dimensional theories obtained by
  dualizing the 3-form field strength, related to the compactification of the
  type~IIA superstring on K3. We call them $\mathcal{N}=2A^{*},d=6$ theories.}
and the family of models with only $n_{T}=n_{V}+1$ tensor multiplets (that we
are going to call $\mathcal{N}=2B,d=6$ theories) give exactly the same family
of models of $\mathcal{N}=2,d=5$ supergravity coupled to $n_{V5}=n_{V}+2$
vector multiplets with a symmetric tensor $C_{IJK}$ with non-vanishing
components $C_{0\, r+1\, s+1}= \tfrac{1}{3!}\eta_{rs}$ with
$r,s=0,\cdots,n_{V}+1$ and $(\eta_{rs})=(+-\cdots-)$.

This situation is analogous to what happens when we dimensionally reduce the
two maximal 10-dimensional supergravities, $\mathcal{N}=2A$ and
$\mathcal{N}=2B$, on a circle and we find the same 9-dimensional maximal
supergravity\footnote{It is unique.} \cite{Bergshoeff:1995as}. In that case,
this coincidence is interpreted as a manifestation at the effective field
theory level of the T-duality existing between the two type~II superstrings
\cite{Dai:1989ua,Dine:1989vu,Witten:1995ex}. The relation between the fields
of the two 10-dimensional supergravities and those of the 9-dimensional one
leads to a direct relation between the 10-dimensional fields of the two
theories: the type~II generalization of the Buscher T-duality rules
\cite{Buscher:1985kb,Buscher:1987sk,Buscher:1987qj} that transform a solution
of one of the 10-dimensional theories admitting one isometry into another
solution of the other theory (also admitting one isometry)
\cite{Bergshoeff:1995as}.

In the present case it is not clear which is the superstring theory associated
to the $\mathcal{N}=2B,d=6$ theories (if any), but the relation we have found
leads to a new generalization of the Buscher rules transforming 6-dimensional
solutions of these theories admitting one isometry (Section~\ref{sec-maps}).

In Section~\ref{sec-applications}, we are going to exploit the
results of Section~\ref{sec-uplifting} to construct new supersymmetric
solutions of the 6-dimensional theories we are discussing
($\mathcal{N}=2A,2A^{*},2B$) by uplifting solutions of the $\mathcal{N}=2,d=5$
theories they all reduced to. We are going to add a new twist to this story,
though. The relations between the fields of two ungauged supergravity theories
related by standard dimensional reduction do not change if we gauge both of
them in the same way. Thus, we can use the uplifting formulae of
Section~\ref{sec-uplifting} to uplift supersymmetric solutions of the same
models of $\mathcal{N}=2,d=5$ supergravity but, now, with non-Abelian
gaugings.

The supersymmetric solutions of general models of gauged $\mathcal{N}=2,d=5$
supergravity were classified in Refs.~\cite{Bellorin:2007yp,Bellorin:2008we},
but the construction of explicit examples in the theories with non-Abelian
gaugings has only been successfully completed recently in
Refs.~\cite{Meessen:2015enl,Ortin:2016bnl}. The method used was essentially
the same we are going to use here: the uplifting of solutions of the
4-dimensional non-Abelian gauged theories which are better understood
\cite{Huebscher:2007hj,Meessen:2008kb,Hubscher:2008yz,Meessen:2015nla,Bueno:2014mea,Bueno:2015wva}. We
are just going to consider the simplest solution in
Ref.~\cite{Meessen:2015enl} to illustrate the procedure, but this will be
enough to produce interesting 6-dimensional solutions.

The uplifting of non-Abelian solutions to the $\mathcal{N}=2A,2A^{*}$ theories
is well justified, but, what is the justification for the $\mathcal{N}=2B$
case if these theories cannot be gauged? We believe that a gauged
$\mathcal{N}=2B,d=6$ theory can be defined at least when the theory is
compactified on a circle. The situation is analogous to that of several
coincident M5-branes which, at least when wrapped on a circle, must be
described by a non-Abelian theory of chiral 2-forms. We do not know how to
write such a theory, but at the massless level, we know it is effectively
described by a standard non-Abelian theory of vector fields in one dimension
less (the theory of coincident D4-branes). We do not know how to describe the
non-Abelian $\mathcal{N}=2B,d=6$ supergravity theory, which only has chiral
2-forms, but we know that, when compactified on a circle, at the massless
level, the theory is described by a standard gauged theory of
$\mathcal{N}=2,d=5$ supergravity with 1-forms as gauge fields. It is in this
limited sense that the non-Abelian solutions of $\mathcal{N}=2B,d=6$
supergravity that we are going to construct should be interpreted.

Finally, Section~\ref{sec-conclusions} contains our conclusions and directions
for future work.

\section{From six to five dimensions}
\label{reduction}

In this section we are going to consider the dimensional reduction of general
theories of ungauged $\mathcal{N}=(2,0),d=6$ supergravity coupled to $n_{T}$
tensor multiplets and $n_{V}$ vector multiplets to five dimensions. We first
review the bosonic sector of the theory explaining our
conventions.\footnote{They are, essentially, those of
  Ref.~\cite{Nishino:1997ff}.} As usual, we denote the 6-dimensional objects
with hats. In particular, $\hat{\mu},\hat{\nu},\ldots =0,\cdots,5$ and
$\hat{a},\hat{b},\ldots=0,\cdots, 5$ are, respectively, 6-dimensional world
and tangent space indices. Our metric has mostly minus signature.

The bosonic fields of the $n_{V}$ vector multiplets, labeled by
$i,j,\ldots=1,\cdots,n_{V}$, are just the 1-form fields $\hat{A}^{i}=
\hat{A}^{i}{}_{\hat{\mu}}d\hat{x}^{\hat{\mu}}$. Their 2-form field strengths
$\hat{F}^{i}=
\tfrac{1}{2}\hat{F}^{i}{}_{\hat{\mu}\hat{\nu}}d\hat{x}^{\hat{\mu}} \wedge
d\hat{x}^{\hat{\nu}}$ are defined as

\begin{equation}
\hat{F}^{i}
\equiv
d\hat{A}^{i}
\,\,\, \Leftrightarrow\,\,\, 
\hat{F}^{i}{}_{\hat{\mu}\hat{\nu}}
\equiv
2\partial_{[\hat{\mu}}\hat{A}^{i}{}_{\hat{\nu}]}\, ,
\end{equation}

\noindent
and are invariant under the gauge transformations

\begin{equation}
\delta \hat{A}^{i}= d\hat{\Lambda}^{i}\, ,
\end{equation}

\noindent
for arbitrary $0$-forms $\hat{\Lambda}^{i}$.

The bosonic fields of the supergravity multiplet are the Sechsbein
$\hat{e}^{\hat{a}}{}_{\hat{\mu}}$, and a 2-form potential $\hat{B}^{0}=
\tfrac{1}{2}\hat{B}^{0}{}_{\hat{\mu}\hat{\nu}}d\hat{x}^{\hat{\mu}} \wedge
d\hat{x}^{\hat{\nu}}$ which satisfies an anti-selfduality constraint whose
explicit form depends on the couplings to the matter fields and will be given
shortly.

The bosonic fields of the $n_{T}$ tensor multiplets, labeled by
$\alpha,\beta,\ldots= 1, \cdots, n_{T}$, are the 2-form potentials
$\hat{B}^{\alpha}{}_{\hat{\mu}\hat{\nu}}$ satisfying selfduality constraints
whose explicit form will also be given shortly, and the real scalars
$\varphi^{\underline{\alpha}}$.  These fields can be seen as coordinates in
the coset space SO$(1,n_{T})/$SO$(n_{T})$. It is convenient to use as coset
representative the SO$(1,n_{T})$ matrix $\hat{L}_{r}{}^{s}$,
$r,s,\ldots=0,1,\cdots,n_{T}$ and it is customary to use the following
notation: $\hat{L}_{r}{}^{s}=(\hat{L}_{r}{},\hat{L}_{r}{}^{\alpha})$ (that
is, $\hat{L}_{r}{}\equiv\hat{L}_{r}{}^{0}$). Then, by definition, these
functions satisfy

\begin{equation}
\eta_{rs}
= 
\eta_{tu}\hat{L}_{r}{}^{t}\hat{L}_{s}{}^{u}
=
\hat{L}_{r}\hat{L}_s-\hat{L}_{r}{}^{\alpha}\hat{L}_{s}{}^{\alpha}\, ,
\hspace{1cm}
\eta_{rs}=\operatorname{diag}(+,-,-,\cdots,-)\, .
\end{equation}

\noindent
Using $\eta_{rs}$ to raise and lower indices we find 

\begin{equation}
\hat{L}^{r}\hat{L}^{s}\eta_{rs}=\hat{L}^{r}\hat{L}_{r}=1\, .
\end{equation}

\noindent
Finally, we define the symmetric SO$(1,n_{T})$ matrix 

\begin{equation}
\mathcal{M}_{rs}
\equiv
\delta_{tu}\hat{L}_{r}{}^{t}\hat{L}_{s}{}^{u}
=
2\hat{L}_{r}\hat{L}_s-\eta_{rs}\, .
\end{equation} 

An SO$(1,n_{T})$-symmetric $\sigma$-model for the scalars
$\varphi^{\underline{\alpha}}$ can be constructed as usual:

\begin{equation}
\hat{L}_{s}{}^{r}\partial_{\hat{a}}\hat{L}^{s}{}_{t} 
\hat{L}_{u}{}^{t}\partial^{\hat{a}}\hat{L}^{u}{}_{r}
=
-\partial_{\hat{a}}\hat{L}^{r} \partial^{\hat{a}}\hat{L}_{r}\, ,
\end{equation}

\noindent
where we have used the above properties of the coset representative. A simple
parametrization of the functions $\hat{L}^{r}$ in terms of the physical
scalars is provided by

\begin{equation}
\hat{L}^{0}
=
(1-\varphi^{\underline{\beta}}\varphi^{\underline{\beta}})^{-1/2}\, ,
\hspace{1cm}
\hat{L}^{\alpha}
=
\varphi^{\underline{\alpha}}
(1-\varphi^{\underline{\beta}}\varphi^{\underline{\beta}})^{-1/2}\, ,
\,\,\,\,
\Rightarrow
\,\,\,\,
\varphi^{\underline{\alpha}} = \hat{L}^{\alpha}/\hat{L}^{0}\, .
\end{equation}

The matter and supergravity 2-forms are combined into a single SO$(1,n_{T})$
vector $(\hat{B}^{r})=(\hat{B}^{0},\hat{B}^{\alpha})$, with 3-form field
strengths $\hat{H}^{r} =
\tfrac{1}{3!}\hat{H}^{r}{}_{\hat{\mu}\hat{\nu}\hat{\rho}}\, d\hat{x}^{\hat{\mu}}
\wedge d\hat{x}^{\hat{\nu}} \wedge d\hat{x}^{\hat{\rho}}$ defined by

\begin{equation}
\hat{H}^{r}
=
d\hat{B}^{r}+\tfrac{1}{2}c^{r}{}_{ij}\hat{F}^{i}\wedge \hat{A}^{j}
\,\,\, \Leftrightarrow \,\,\,
\hat{H}^{r}{}_{\hat{\mu}\hat{\nu}\hat{\rho}}
=
3\partial_{[\hat{\mu}}\hat{B}^{r}{}_{\hat{\nu}\hat{\rho}]}
+\tfrac{3}{2}c^{r}{}_{ij}\hat{F}^{i}{}_{[\hat{\mu}\hat{\nu}}
\hat{A}^{i}{}_{\hat{\rho}]}\, ,
\end{equation}

\noindent
where $c^{r}{}_{ij}$ is an array of constant positive-definite matrices. They
are invariant under the gauge transformations

\begin{equation}
\delta \hat{B}^{r}
=
d\hat{\chi}^{r}-\tfrac{1}{2}c^{r}{}_{ij}\hat{F}^{i}\hat{\Lambda}^{j},
\end{equation}

\noindent
for arbitrary 1-forms $\hat{\chi}^{r}$, and they are constrained to satisfy
the (anti-) selfduality constraint

\begin{equation}
\label{duality}
\mathcal{M}_{rs}\hat{H}^{s}=-\eta_{rs}\star\hat{H}^{s}\, ,
\,\,\,\,
\mbox{where}
\,\,\,\, 
\eta_{rs}=\operatorname{diag}(+,-,-,\cdots,-)\, .
\end{equation}

Using this constraint in the Bianchi identity of the 3-form field strengths

\begin{equation}
d\hat{H}^{r} -\tfrac{1}{2}c^{r}{}_{ij}\hat{F}^{i}\wedge \hat{F}^{j}=0\, ,
\end{equation}

\noindent
one obtains the equation of motion of the 2-forms:

\begin{equation}
\label{eq:Hreom}
d\left(\mathcal{M}_{rs}\star\hat{H}^{s}\right)
+\tfrac{1}{2}c_{r\, ij}\hat{F}^{i}\wedge \hat{F}^{j}=0\, .
\end{equation}

It is convenient to work with the action of the theory but, in general, these
theories do not have a covariant action, due to (anti-) selfduality
constraints satisfied by the 3-forms \cite{Marcus:1982yu}. Nevertheless,
sometimes, it is possible to construct \textit{pseudoactions}
\cite{Bergshoeff:1995sq} which give the correct equations of motion of the
theory upon use of the (anti-) selfduality constraints in the Euler-Lagrange
equations that follow from them. The action of the dimensionally reduced
theory can then be derived by following these directions:

\begin{enumerate}
\item Dimensionally reduce the pseudoaction and the (anti-) selfduality
  constraints in the standard way.
\item Poincar\'e-dualize the highest-rank potentials arising from the (anti-)
  selfdual potentials in the dimensionally-reduced pseudoaction.
\item Identify the resulting potentials with the lowest-rank potentials arising
  from the (anti-) selfdual potentials. This identification should be
  completely equivalent to the use of the dimensionally reduced (anti-)
  selfduality constraint in the action.
\end{enumerate}

A well-known example of this procedure is the dimensional reduction to $d=9$
of the $\mathcal{N}=2B,d=10$ supergravity theory
\cite{Schwarz:1983wa,Schwarz:1983qr,Howe:1983sra} carried out in
Ref.~\cite{Meessen:1998qm}: in this case there is a RR 4-form potential
$\hat{C}^{(4)}$ whose 5-form field strength $\hat{G}^{(5)}$ is self-dual
$\hat{G}^{(5)}=\star_{10} \hat{G}^{(5)}$ and the equations of motion can be
derived from the pseudoaction constructed in Ref.~\cite{Bergshoeff:1995sq} by
imposing a selfduality constraint. The dimensional reduction of the 4-form
potential $\hat{C}^{(4)}$ gives rise to a 4- and a 3-form $C^{(4)},C^{(3)}$
potentials whose 5- and 4-form field strengths $G^{(5)}$ and $G^{(4)}$ are
related by the dimensionally reduced selfduality constraint $G^{(5)}\sim \star
G^{(4)}$. Following the above recipe, in Ref.~\cite{Meessen:1998qm} the
pseudoaction and selfduality constraint were reduced to $d=9$ first. Then, the
9-dimensional 4-form potential $C^{(4)}$ was Poincar\'e-dualized into a
9-dimensional 3-form potential $\tilde{C}^{(3)}$ in the pseudoaction. At this
point the theory has two different 3-form potentials $\tilde{C}^{(3)}$ and
$C^{(3)}$ and the selfduality constraint takes the form
$\tilde{G}^{(4)}=G^{(4)}$ indicating that the two 3-forms are one and the same
$\tilde{C}^{(3)}=C^{(3)}$. Making this identification in the pseudoaction
gives the correct 9-dimensional action.

In the case at hands, the bosonic equations of motion (in particular,
Eq.~(\ref{eq:Hreom})) can be found by varying the pseudoaction

\begin{equation}
\label{pseudoaction}
\hat{S}
=
\int d^{6}\hat{x}\sqrt{|\hat{g}|}
\Big\{
\hat{R} 
-\partial_{\hat{a}}\hat{L}^{r}\partial^{\hat{a}}\hat{L}_{r}
+\tfrac{1}{3}\mathcal{M}_{rs}\hat{H}^{r}{}_{\hat{a}\hat{b}\hat{c}}
\hat{H}^{s\, \hat{a}\hat{b}\hat{c}}
-\hat{L}_{r}c^{r}{}_{ij}\hat{F}^{i}{}_{\hat{a}\hat{b}}\hat{F}^{j\, \hat{a}\hat{b}}
-\tfrac{1}{4}c_{r\, ij}\hat{\epsilon}^{\hat{a}\hat{b}\hat{c}\hat{d}\hat{e}\hat{f}}
\hat{B}^{r}{}_{\hat{a}\hat{b}}\hat{F}^{i}{}_{\hat{c}\hat{d}}
\hat{F}^{j}{}_{\hat{e}\hat{f}}
\Big\}\, .
\end{equation}

\noindent
and imposing on the resulting Euler-Lagrange equations the (anti-) selfduality
conditions Eqs.~(\ref{duality}).  However, due to the Chern-Simons term, this
action is gauge invariant if and only if the following condition holds
\cite{Hartong:2009vc}

\begin{equation}
\label{gaugecond}
\eta_{rs}c^{r}{}_{i(j}c^{s}_{\ kl)}=0\, ,
\end{equation}

\noindent
and we will assume this condition to hold through our work. Only then one
gets consistent five-dimensional theories.

\subsection{Reduction of the fields}

Having described the bosonic sector of the theories we want to study, we are
now ready to reduce them to $d=5$. 

We are going to follow the standard procedure proposed in
Ref.~\cite{Scherk:1979zr} with the particular conventions of
Ref.~\cite{Ortin:2015hya}. Thus, we assume that none of the fields depends
explicitly on the compact coordinate, that we will call $z$, we split the
world and tangent-space indices as follows

\begin{equation}
\hat{\mu}=\mu,\underline z\, , \hspace{1cm} \hat{a}=a,z\, ,
\end{equation}

\noindent 
and we decompose the components of the Sechsbein basis (which we choose to be
upper-triangular) $\hat{e}^{\hat{a}}{}_{\hat{\mu}}$ into those of a F\"unfbein
$e^{a}{}_{\mu}$, a (Kaluza-Klein (KK)) vector $A_{\mu}$ and a KK scalar $k$ as
follows:

\begin{equation}
\left(\hat{e}^{\hat{a}}{}_{\hat{\mu}} \right)
=
\begin{pmatrix}
e^{a}{}_{\mu} & kA_{\mu}\\
0 & k \\
\end{pmatrix}\, , 
\hspace{1.5cm}
\left( \hat{e}_{\hat{a}}{}^{\hat{\mu}} \right)
=
\begin{pmatrix}
e_{a}{}^{\mu} & -A_{a} \\
0 & k^{-1}\\
\end{pmatrix}\, ,
\end{equation}

\noindent
where $A_{a}= e_{a}{}^{\mu}A_{\mu}$.

The scalars are the same $z$-independent functions in both dimensions. In
particular, $\hat{L}_{r}=L_{r}$.

The vector fields $\hat{A}^{i}$ decompose into vector fields $A^{i}$ and
scalar fields $l^{i}$ as follows:

\begin{eqnarray}
\hat{A}^{i}{}_{a}
\equiv 
A^{i}{}_{a}\,\,\, 
& \Leftrightarrow & \,\,\,
\hat{A}^{i}{}_{\mu}=A^{i}{}_{\mu}+l^{i}A_{\mu}\, ,
\\
& & \nonumber \\
\hat{A}^{i}{}_{z}
\equiv 
k^{-1}l^{i} 
& \Leftrightarrow & 
\hat{A}^{i}{}_{\underline{z}}=l^{i}\, .
\end{eqnarray}

\noindent
This leads to the following decomposition of the vector field strengths:

\begin{eqnarray}
\label{Ficurly}
\hat{F}^{i}{}_{ab} 
& = & 
\mathcal{F}^{i}{}_{ab}  
=
F^{i}{}_{ab}+l^{i}F_{ab}\, ,
\\
& & \nonumber \\
\hat{F}^{i}{}_{az}
& = &
k^{-1}\partial_{a}l^{i}\, ,
\end{eqnarray}
 
\noindent
where $F^{i}$ and $F$ are the 5-dimensional field strengths

\begin{equation}
F^{i}\equiv dA^{i}\, ,
\hspace{1cm}
F \equiv dA\, .
\end{equation}

Each 2-form $\hat{B}^{r}$ produces a 2- and 1-form in five dimensions
($B^{r}$ and $A^{r}$ respectively). They will be related by the (anti-)
selfduality constraints. It turns out that the following definitions give
potentials with good gauge transformation properties:

\begin{eqnarray}
\hat{B}^{r}{}_{\mu\underline{z}}
& \equiv & 
A^{r}{}_{\mu}+\tfrac{1}{2}c^{r}{}_{ij}l^{i}A^{j}{}_{\mu}\, , \\
& & \nonumber \\
\hat{B}^{r}{}_{\mu\nu}
& \equiv & 
B^{r}{}_{\mu\nu}-A_{[\mu}A^{r}{}_{\nu]}
-c^{r}{}_{ij}A_{[\mu}A^{i}{}_{\nu]}l^{j}\, .
\end{eqnarray}

The 3-form field strengths $\hat{H}^{r}$ decompose as follows:

\begin{eqnarray}
\hat{H}^{r}{}_{abc} 
& \equiv & 
H^{r}{}_{abc}\, ,  
\\
& & \nonumber \\
\label{Frcurly}
\hat{H}^{r}{}_{abz}
& \equiv & 
k^{-1}\mathcal{F}^{r}{}_{ab}
\equiv 
k^{-1}
\left[ 
F^{r}+c^{r}{}_{ij}l^{i}F^{j}+\tfrac{1}{2}c^{r}{}_{ij}l^{i}l^{j}F
\right]
\, ,
\end{eqnarray}

\noindent
where 

\begin{eqnarray}
H^{r}
& = & 
dB^{r}-\tfrac{1}{2}F\wedge A^{r}-\tfrac{1}{2}F^{r}\wedge A
+\tfrac{1}{2}c^{r}{}_{ij}F^{i}\wedge A^{j}\, ,
\\
& & \nonumber \\
F^{r}
& = & 
dA^{r}\,.
\end{eqnarray}

This completely fixes the reduction of fields and field strengths. 
Plugging these decompositions in the pseudoaction Eq.~(\ref{pseudoaction}) 
together with the decomposition of the Levi-Civita symbol 
\begin{equation}
\hat{\epsilon}^{abcdez} \equiv \epsilon^{abcde}\, ,
\end{equation}

\noindent
we get in a straightforward manner the 5-dimensional pseudoaction

\begin{equation}
\begin{aligned}
S=\int d^{5}x\sqrt{|g|}k\Big\{
&
R-\tfrac{1}{4}k^{2}F^{2}-\partial_{\mu}L^{r}\partial^{\mu}L_{r}
+2k^{-2}L_{r}c^{r}{}_{ij}\partial_{\mu}l^{i}\partial^{\mu}l^{j}
\\
&
\\
&
+\tfrac{1}{3}\mathcal{M}_{rs}H^{r}H^{s}
-k^{-2}\mathcal{M}_{rs}\mathcal{F}^{r}\mathcal{F}^{s}
-L_{r}c^{r}{}_{ij}\mathcal{F}^{i}\mathcal{F}^{j}
\\
&
\\
&
+\frac{k^{-1}\epsilon}{6\sqrt{|g|}}c_{r\, ij} 
\left[
H^{r}(\mathcal{F}^{i}l^{j}-2\partial l^{i}A^{j})
-3\mathcal{F}^{r}\mathcal{F}^{i}A^{j}
\right]
\Big\}\, ,
\end{aligned}
\end{equation}

\noindent
where the indices are assumed to be contracted in the obvious way:
$\mathcal{F}^{r}\mathcal{F}^{s}\equiv\mathcal{F}^{r}{}_{\mu\nu}\mathcal{F}^{s\
  \mu\nu}$, $\epsilon H^{r}c_{r\, ij}(\mathcal{F}^{i}l^{j}-2\partial
l^{i}A^{j})=\epsilon^{\mu\nu\rho\kappa\sigma}H^{r}{}_{\mu\nu\rho}c_{r\,
  ij}(\mathcal{F}^{i}{}_{\kappa\sigma}l^{j}
-2\partial_{[\kappa}l^{i}A^{j}{}_{\sigma]})$, etc.

Finally, we make a rescaling of the metric in order to express the action in
the ``Einstein frame'' metric $g_{E\, \mu\nu}$ (minimal coupling to Ricci
scalar) in the following way:

\begin{equation}
g_{\mu\nu}=k^{-2/3}g_{E\, \mu\nu}\, ,
\end{equation}

\noindent
and redefine the KK scalar $k$ in order to give it a kinetic term with
standard normalization

\begin{equation}
k=e^{\sqrt{3/8}\phi}\, .
\end{equation}

The result, up to total derivatives, is the pseudoaction

\begin{equation}
\label{pseudoaction2}
\begin{aligned}
S =\int d^{5}x\sqrt{|g_{E}|}\Big\{
&
R_{E}+\tfrac{1}{2}(\partial\phi)^{2}
-\partial_{\mu}L^{r}\partial^{\mu}L_{r}
+2e^{-\sqrt{3/2}\phi}L_{r}c^{r}{}_{ij}\partial_{\mu}l^{i}\partial^{\mu}l^{j}
-\tfrac{1}{4}e^{\sqrt{8/3}\phi}F^{2}
\\
&
\\
&
-e^{-\sqrt{2/3}\phi}\mathcal{M}_{rs}\mathcal{F}^{r}\mathcal{F}^{s}
-L_{r}c^{r}{}_{ij}e^{\phi/\sqrt{6}}\mathcal{F}^{i}\mathcal{F}^{j} 
+\tfrac{1}{3}e^{\sqrt{2/3}\phi}\mathcal{M}_{rs}H^{r}H^{s}
\\
&
\\
&
+\frac{\epsilon}{6\sqrt{|g_{E}|}}c_{r\, ij}
\left[
H^{r}(\mathcal{F}^{i}l^{j}-2\partial l^{i}A^{j})
-3\mathcal{F}^{r}\mathcal{F}^{i}A^{j}
\right]\Big\}\, .
\\
\end{aligned}
\end{equation}

The reduction of the (anti-) selfduality constraints Eqs.~(\ref{duality})
offers no problems and becomes a duality relation between the 2- and 1-form
potentials $B^{r},A^{r}$

\begin{equation}
\label{duality2}
\mathcal{M}_{rs}H^{s}=-e^{-\sqrt{2/3}\phi}\eta_{rs}\star\mathcal{F}^{s}\, .
\end{equation}

The equations of motion of the 5-dimensional theory can be obtained by varying
the above pseudoaction and imposing the duality constraints. However, in
order to identify the 5-dimensional theories obtained with models of
$\mathcal{N}=2,d=5$ supergravity coupled to vector multiplets it is convenient
to eliminate this constraint. We carry out this task next.

\subsection{Dualization}

Following the procedure outlined at the beginning of this section, we are
going to Poincar\'e dualize the 2-forms $B^{r}$ into 1-forms
$\tilde{A}_{r}$. First, we are going replace the 2-forms $B^{r}$ by their
3-form field strengths $H^{r}$ as variables of the pseudoaction
Eq.~(\ref{pseudoaction2}). This is possible because the pseudoaction only
depends on the 2-forms through their field strengths. However, we have to add
a Lagrange-multiplier term to enforce the Bianchi identities of the $H^{r}$,
which have the form

\begin{equation}
4\partial_{[\mu}H^{r}{}_{\nu\rho\sigma]}+6F^{r}{}_{[\mu\nu}F_{\rho\sigma]}
-3c^{r}{}_{ij}F^{i}{}_{[\mu\nu}F^{j}{}_{\rho\sigma]}=0\, .
\end{equation}

The Lagrange-multiplier term to be added to the pseudoaction to enforce the
Bianchi identity is (again, with the indices contracted in the obvious way)

\begin{equation}
\frac{\epsilon}{\sqrt{|g_{E}|}}\tilde{A}_{r}
\Big(
\partial H^{r}+\tfrac{3}{2}F^{r}F-\tfrac{3}{4}c^{r}{}_{ij}F^{i}F^{j}
\Big)\, ,
\end{equation}

\noindent
where the Lagrange multiplier is the 1-form field $\tilde{A}_{r}$. 

Adding this term to the pseudoaction and integrating it by parts we get

\begin{equation}
\begin{aligned}
S=\int d^{5}x\sqrt{|g_{E}|}\Big\{
&
R_{E}+\tfrac{1}{2}(\partial\phi)^{2}
-\partial_{\mu}L^{r}\partial^{\mu}L_{r}
+2e^{-\sqrt{3/2}\phi}L_{r}c^{r}{}_{ij}\partial_{\mu}l^{i}\partial^{\mu}l^{j}
-\tfrac{1}{4}e^{\sqrt{8/3}\phi}F^{2}
\\
&
\\
&
-e^{-\sqrt{2/3}\phi}\mathcal{M}_{rs}\mathcal{F}^{r}\mathcal{F}^{s}
-L_{r}c^{r}{}_{ij}e^{\phi/\sqrt{6}}\mathcal{F}^{i}\mathcal{F}^{j}
+\tfrac{1}{3}e^{\sqrt{2/3}\phi}\mathcal{M}_{rs}H^{r}H^{s}
\\
&
\\
&
+\frac{\epsilon}{6\sqrt{|g_{E}|}}
\left[
c_{r\, ij}H^{r}(\mathcal{F}^{i}l^{j}-2\partial l^{i}A^{j})
-3c_{r\, ij}\mathcal{F}^{r}\mathcal{F}^{i}A^{j}
\right.
\\
&
\\
&
\left.
+3\tilde{F}_{r}(H^{r}+\tfrac{3}{2}FA^{r}+\tfrac{3}{2}F^{r}A
-\tfrac{3}{2}c^{r}{}_{ij}F^{i}A^{j})\right]
\Big\},
\end{aligned}
\end{equation}

\noindent
where

\begin{equation}
\tilde{F}_{r} \equiv d\tilde{A}_{r}\, .  
\end{equation}

Since in this pseudoaction $H^{r}$ is an independent field, we can compute its
field equation, which will relate it to $\tilde{F}_{r}$. It is given by

\begin{equation}
\label{dualization}
\mathcal{M}_{rs}H^{s}
=
-\tfrac{1}{2}e^{-\sqrt{2/3}\phi}
\star\Big[c_{r\, ij}(\mathcal{F}^{i}l^{j}-2\partial l^{i}A^{j})
+3\tilde{F}_{r}\Big]\, ,
\end{equation}

\noindent
This equation can be used to eliminate completely $H^{r}$ from the
pseudoaction and from the duality relation Eq.~(\ref{duality2}). After this
operation, the 2-forms $B^{r}$ have disappeared from both, having been replaced
by the dual 1-forms $\tilde{A}_{r}$. We only write explicitly the constraint
after this replacement (and some massaging):

\begin{equation}
\tilde{F}_{r}
=
\tfrac{2}{3}\Big(\eta_{rs}F^{s}+c_{r\, ij}\partial(l^{i}A^{j})\Big)\, , 
\end{equation}

\noindent
which implies the following algebraic relation between potentials

\begin{equation}
\tilde{A}_{r}
=
\tfrac{2}{3}\eta_{rs}A^{s}+\tfrac{1}{3}c_{r\, ij}l^{i}A^{j}\, ,
\end{equation}

\noindent
that we can use in the pseudoaction to eliminate completely $\tilde{A}_{r}$.
After this operation  the 1-forms $A^{r}$ are the only fields remaining from
the reduction of the 2-forms $B^{r}$. Furthermore, there are no constraints to
be imposed and the pseudoaction is the standard action

\begin{equation}
\begin{aligned}
S=\int d^{5}x\sqrt{|g_{E}|}\Bigg\{
&
R_{E}+\tfrac{1}{2}(\partial\phi)^{2}-\partial_{\mu}L^{r}\partial^{\mu}L_{r}+2e^{-\sqrt{3/2}\phi}L_{r}c^{r}{}_{ij}\partial_{\mu}l^{i}\partial^{\mu}l^{j}
\\
&
\\
&
-\tfrac{1}{4}e^{\sqrt{8/3}\phi}F^{2}
-2e^{-\sqrt{2/3}\phi}\mathcal{M}_{rs}\mathcal{F}^{r}\mathcal{F}^{s}
-L_{r}c^{r}{}_{ij}e^{\phi/\sqrt{6}}\mathcal{F}^{i}\mathcal{F}^{j} 
\\
&
\\
&
+\frac{\epsilon}{\sqrt{|g_{E}|}}\Big(\eta_{rs}F^{r}F^{s}A
-c_{r\, ij}F^{i}F^{j}A^{r}\Big)\Bigg\}\, .
\end{aligned}
\end{equation}

\subsection{Identification with five-dimensional supergravity}

The next step is to identify the previous theory as a model of
$\mathcal{N}=1,d=5$ supergravity coupled to $n_{V5}$ vector multiplets. These
theories\footnote{We use the conventions of Refs.~\cite{Bergshoeff:2004kh} and
  \cite{Bellorin:2006yr}.} contain $n_{V5}+1$ 1-form fields $A^{I}$,
$I,J,\ldots =0,1,\cdots,n_{V5}$ and $n_{V5}$ scalars $\phi^{x}$, $x,y,\ldots =
1,\cdots,n_{V5}$, and their interactions (in fact, the whole theory) are
determined by the constant and completely symmetric tensor $C_{IJK}$. In
particular, the scalar manifold is the $n_{V5}$-dimensional hypersurface in
$\mathbb{R}^{n_{V5}+1}$ defined by the cubic equation

\begin{equation}
\label{eq:Ch3=1}
C_{IJK}h^{I}(\phi)h^{J}(\phi)h^{K}(\phi)=1,
\end{equation}

\noindent 
the kinetic matrix of the vector fields $a_{IJ}(\phi)$ is given by

\begin{equation}
\label{aIJ}
a_{IJ}
=
-2C_{IJK}h^{K} +3h_{I}h_{J}\, ,  
\end{equation}

\noindent
where the $h_{I}(\phi)$ are defined by

\begin{equation}
\label{eq:h_I}
h_{I}\equiv C_{IJK}h^{J}h^{K}\, ,
\end{equation}

\noindent
and the $\sigma$-model metric $g_{xy}(\phi)$ is given by 

\begin{equation}
\label{gxy}
g_{xy}
\equiv
3a_{IJ}\frac{\partial h^{I}}{\partial\phi^{x}}
\frac{\partial h^{J}}{\partial\phi^{y}}
=
-2C_{IJK}\frac{\partial h^{I}}{\partial\phi^{x}}
\frac{\partial h^{J}}{\partial\phi^{y}} h^{K}\, .
\end{equation}

The action is given by 

\begin{equation}
\label{5dsugraaction}
S
=
\int d^{5}x\sqrt{|g|}\Bigg\{
R+\tfrac{1}{2}g_{xy}\partial_{\mu}\phi^{x}\partial^{\mu}\phi^{y}
-\tfrac{1}{4}a_{IJ}F^{I}F^{J}
+\frac{\epsilon}{12\sqrt{3}\sqrt{|g|}}C_{IJK}F^{I}F^{J}A^{K}
\Bigg\}\, .
\end{equation}

In order to identify the models corresponding to the theories we have obtained
by dimensional reduction, we start by rescaling the vector fields 

\begin{equation}
A\rightarrow \tfrac{1}{\sqrt{12}}A\, ,
\hspace{1cm}
A^{r}\rightarrow \tfrac{1}{\sqrt{12}}A^{r}\, , 
\hspace{1cm}
A^{i}\rightarrow \tfrac{1}{\sqrt{12}}A^{i}\, ,
\end{equation}

\noindent
so that the action becomes

\begin{equation}
\label{reducedaction}
\begin{aligned}
S = \int d^{5}x\sqrt{|g_{E}|}\Bigg\{
&
R_{E}+\tfrac{1}{2}(\partial\phi)^{2}
-\partial_{\mu}L^{r}\partial^{\mu}L_{r}
+2e^{-\sqrt{3/2}\phi}L_{r}c^{r}{}_{ij}\partial_{\mu}l^{i}\partial^{\mu}l^{j}
\\
&
\\
&
-\tfrac{1}{48}e^{\sqrt{8/3}\phi}F^{2}
-\tfrac{1}{12}L_{r}c^{r}{}_{ij}e^{\phi/\sqrt{6}}(F^{i}{}_{\mu\nu}+l^{i}F_{\mu\nu})(F^{j}{}_{\mu\nu}+l^{j}F_{\mu\nu})
\\
&
\\
&
-\tfrac{1}{6}e^{-\sqrt{2/3}\phi}\mathcal{M}_{rs}
\Big(F^{r}{}_{\mu\nu}+c^{r}{}_{ij}l^{i}F^{j}{}_{\mu\nu}
+\tfrac{1}{2}c^{r}{}_{ij}l^{i}l^{j}F_{\mu\nu}\Big)\Big(F^{s}{}_{\mu\nu}
+c^{s}{}_{ij}l^{i}F^{j}{}_{\mu\nu}
+\tfrac{1}{2}c^{s}{}_{ij}l^{i}l^{j}F_{\mu\nu}\Big) 
\\
&
\\
&
+\frac{\epsilon}{12\sqrt{3}\sqrt{|g_{E}|}}
\Big(\tfrac{1}{2}\eta_{rs}F^{r}F^{s}A
-\tfrac{1}{2}c_{r\, ij}F^{i}F^{j}A^{r}\Big)\Bigg\}\, .
\end{aligned}
\end{equation}

Comparing this theory with Eq.~(\ref{5dsugraaction}) we first see that
$n_{V5}=n_{T}+n_{V}+1$ (there is a total of $n_{T}+n_{V}+2$ 1-forms). We can
decompose the 5-dimensional index $I$ as $I=0,r+1,i+n_{T}+1$ where the indices
take the values $r=0,\ldots, n_{T}$, $i=1,\ldots ,n_{V}$ and identify

\begin{equation}
  A^{0}=A\, , \hspace{1cm} A^{I=r+1}=A^{r}\, ,  \hspace{1cm} A^{I=i+n_{T}+1}=A^{i}\, ,
\end{equation}

\noindent
where the fields in the l.h.s.'s are those of Eq.~(\ref{5dsugraaction}) and
the fields in the r.h.s.'s are those of Eq.~(\ref{reducedaction}).

We can also identify the components of the $C_{IJK}$ tensor that characterizes
the model of $\mathcal{N}=2,d=5$ supergravity

\begin{equation}
\label{CIJK}
C_{0\, r+1\, s+1} = \tfrac{1}{3!}\eta_{rs}\, , 
\hspace{1cm} 
C_{r+1\, i+n_{T}+1\, j+n_{T}+1}=-\tfrac{1}{3!}c_{r\, ij}\, .
\end{equation}

\noindent
We will discuss later the properties of these models, picking two particular
subfamilies. Now, knowing $C_{IJK}$ and the expected forms of $a_{IJ}$ and
$g_{xy}$, we can identify the scalar fields of Eq.~(\ref{reducedaction}) with
the scalar functions $h^{I}$ and the physical scalars $\phi^{x}$. 

The components of $a_{IJ}$ in Eq.~(\ref{reducedaction}) are 

\begin{equation}
\begin{array}{rcl}
a_{00} & = & 
\tfrac{1}{12}\Big[e^{2\phi/\sqrt{6}}+2L_{r}\xi^{r}e^{-\phi/\sqrt{6}}\Big]^{2}\,
,
\\
& & \\
a_{0\, r+1} 
& = & 
\tfrac{1}{3}\mathcal{M}_{rs}\xi^{s}e^{-\sqrt{2/3}\phi},
\\
& & \\
a_{0\, i+n_{T}+1} 
& = & 
\tfrac{1}{3}L_{r}c^{r}{}_{ij}l^{j}e^{-\phi/\sqrt{6}}\Big(e^{2\phi/\sqrt{6}}+2L_s\xi^{s}e^{-\phi/\sqrt{6}}\Big),\\
& & \\
a_{r+1\, s+1} 
& = & 
\tfrac{2}{3}e^{-\sqrt{2/3}\phi}\mathcal{M}_{rs},
\\
& & \\
a_{r+1\, i+n_{T}+1} 
& = & 
\tfrac{2}{3}e^{-\sqrt{2/3}\phi} \mathcal{M}_{rs} c^{s}{}_{ij}l^{j},
\\
& & \\
a_{i+n_{T}+1\, j+n_{T}+1} 
& = & 
\tfrac{2}{3}e^{-\sqrt{2/3}\phi}\mathcal{M}_{rs}
c^{r}{}_{ik}c^{s}{}_{jl}l^kl^l+\tfrac{1}{3}e^{\phi/\sqrt{6}}L_{r}c^{r}{}_{ij},
\end{array}
\end{equation}

\noindent
where $\xi^{r}\equiv c^{r}{}_{ij}l^{i}l^{j}$ and we have made some
simplifications by using the properties $L^{r}L_{r}=1$, $\xi^{r}\xi_{r}=0$,
$\xi^{r}c_{r\, ij}l^{i}=0$ and
$\mathcal{M}_{rs}=2L_{r}L_s-\eta_{rs}$. Finally, if we use as physical scalar
fields $(\phi^{x})=(\phi^{1},\cdots,\phi^{n_{V}+n_{T}+1})
=(\phi,\varphi^{\underline{\alpha}},l^{i})$, we see from
(\ref{reducedaction}) that only the diagonal components of $g_{xy}$ are
non-vanishing:

\begin{equation}
\begin{array}{rcl}
g_{11} 
& = & 
1\, ,
\\
& & \\
g_{\underline{\alpha}+1\, \underline{\beta}+1} 
& = & 
-2\partial_{\underline{\alpha}}L^{r}\partial_{\underline{\beta}}L_{r},
\\
& & \\
g_{i+n_{T}+1\, j+n_{T}+1} 
& = & 
4e^{-\sqrt{3/2}\phi}L_{r}c^{r}{}_{ij}\, .\\
\end{array}
\end{equation}

Comparing these expressions with the formulae Eqs.~(\ref{aIJ}) and (\ref{gxy})
for the theories with symmetric tensor given by Eq.~(\ref{CIJK}) we conclude
that the scalar functions $h^{I}$ are given by 

\begin{equation}
h^{0}
=
2e^{-2\phi/\sqrt{6}}\, , 
\hspace{1cm} 
h^{r}
=
L^{r}e^{\phi/\sqrt{6}}+\xi^{r}e^{-2\phi/\sqrt{6}}\, ,
\hspace{1cm} 
h^{i}
=
-2e^{-2\phi/\sqrt{6}}l^{i}\, .
\end{equation}

For the sake of convenience we also give the $h_{I}$:

\begin{equation}
h_{0}
= 
\tfrac{1}{6}\Big(e^{2\phi/\sqrt{6}}+2\xi_{r}L^{r}e^{-\phi/\sqrt{6}}\Big)\, ,
\hspace{.6cm} 
h_{r}
=
\tfrac{2}{3}L_{r}e^{-\phi/\sqrt{6}}\, ,
\hspace{.6cm} 
h_{i}
=
\tfrac{2}{3}e^{-\phi/\sqrt{6}}c_{r\, ij}L^{r}l^{j}\, .
\end{equation}

We are interested in two particular cases which correspond to models of the
same family with all the scalars in symmetric spaces SO$(1,n)/$SO$(n)$ for
some value of $n$: $n_{V}=0$, which has SO$(1,n_{T}+1)/$SO$(n_{T}+1)$ and
$n_{T}=1$, which has SO$(1,n_{V}+2)/$SO$(n_{V}+2)$. Let us review them more
closely.

\subsubsection{Case $n_{V}=0$}

If we begin with a six-dimensional theory with an arbitrary number $n_{T}$ of
tensor multiplets and no vector multiplets, we arrive to the model with
$n_{V5}=n_{T}+1$ characterized by

\begin{equation}
C_{0rs}=\tfrac{1}{3!}\eta_{rs}\, ,
\end{equation} 

\noindent
and with the parametrization
 
\begin{equation}
h^{0}=2e^{-2\phi^{1}/\sqrt{6}}\, , 
\hspace{1cm} 
h^{r}=e^{\phi^{1}/\sqrt{6}}L^{r}\, ,
\end{equation}

\noindent
with $L^{r}=L^{r}(\phi^{2},\cdots ,\phi^{n_{T}+1})$. 

The $n_{V5}=n_{T}+1$ scalars of these models parametrize the coset
SO$(1,n_{T}+1)/$SO$(n_{T}+1)$. Upon dimensional reduction one obtains an
$ST[2,n_{T}+1]$ model of $\mathcal{N}=2,d=4$ supergravity coupled to
$n_{V4}=n_{V5}+1=n_{T}+2$ vector multiplets parametrizing the coset space
$\frac{\mathrm{SL}(2,\mathbb{R})}{\mathrm{SO}(2)} \times 
\frac{\mathrm{SO}(2,n_{T}+1)}{\mathrm{SO}(2)\times \mathrm{SO}(n_{T}+1)}$.

\subsubsection{Case $n_{T}=1$}

Let us start from a six-dimensional theory with $n_{T}=1$ and an arbitrary
number of vector multiplets $n_{V}$ and let us choose the coefficients $c_{r\,
  ij}$ to be

\begin{equation}
c_{0\, ij}=c_{1\, ij}=\delta_{ij}\, ,
\end{equation}

\noindent
which is a particularly simple solution of the constraint
Eq.~(\ref{gaugecond}). These theories contain two 2-forms of opposite
selfduality that can be combined into a single, unconstrained, 2-form that can
be identified with the Kalb-Ramond field, a single scalar that can be
identified with the dilaton field and a set of Abelian vector fields. These
theories can be obtained by toroidal compactification to 6 dimensions and
subsequent truncation of the Heterotic String theory, assuming that the number
of Abelian vectors does not exceed 16. We will show later how to rewrite it in
the standard form. Now we just want to show that, after dimensional reduction,
these theories also belong to the same family as those of the $n_{V}=0$ case.

With the above choice of coefficients, the parametrization of $\tilde h^{i}$
is given by\footnote{We are going to denote the objects of these theories with
  tildes.}

\begin{equation}
\label{parametrization2}
\begin{array}{rclrcl}
\tilde{h}^{0}
& = & 
2e^{-2 \phi/\sqrt{6}}\, , 
& 
\tilde{h}^{1}
& = & 
\tilde L^{0}e^{\phi/\sqrt{6}}+l^{2}e^{-2\phi/\sqrt{6}}\, ,
\\
& & & & & \\
\tilde{h}^{2}
& = & 
\tilde L^{1}e^{\phi/\sqrt{6}}-l^{2}e^{-2\phi/\sqrt{6}}\, ,
\hspace{1cm}
& 
\tilde{h}^{i}
& = & 
-2e^{-2\phi/\sqrt{6}}l^{i}\, .
\end{array}
\end{equation}

These functions satisfy the equation

\begin{equation}
1 = \tilde C_{IJK}\tilde{h}^{I}\tilde{h}^{J}\tilde{h}^{K}
=
\tfrac{1}{2}\tilde{h}^{0}\big[(\tilde{h}^{1})^{2}-(\tilde{h}^{2})^{2}\big]
-\tfrac{1}{2}(\tilde{h}^{1}+\tilde{h}^{2})\tilde{h}^{i}\tilde{h}^{i}\, .
\end{equation}

However, we are free to make linear transformations of the $\tilde{h}^{I}$ and
$A^{I}$ in order to obtain equivalent theories. In particular, if we perform
the transformation $(\tilde{h}^{0},
\tilde{h}^{1},\tilde{h}^{2},\tilde{h}^{i})\rightarrow (h^{0}, h^{r})$, with
$r=1,2,i+2$, given by

\begin{equation}
\label{lineartransform}
\begin{aligned}
\tilde{h}^{0}&=h^{1}+h^{2},\\
\tilde{h}^{1}&=\tfrac{1}{2}(h^{0}+h^{1}-h^{2}),\\
\tilde{h}^{2}&=\tfrac{1}{2}(h^{0}-h^{1}+h^{2}),\\
\tilde{h}^{i}&=h^{i+2}\, ,
\end{aligned}
\end{equation}

\noindent
we find that the new variables satisfy

\begin{equation}
1 = \tfrac{1}{2}h^{0}\big((h^{1})^{2}-(h^{2})^{2}-h^{i+2}h^{i+2}\big)
=\tfrac{1}{2}h^{0}h^{r}h^{s}\eta_{rs} \equiv  C_{IJK}h^{i}h^{j}h^{K}\, ,
\end{equation}

\noindent
so these models are equivalent to those with
$C_{0rs}=\frac{1}{3!}\eta_{rs}$. 

We conclude that $\mathcal{N}=(2,0),d=6$ supergravity coupled to $n_{T}$
tensor multiplets gives the same five-dimensional supergravity model as
$\mathcal{N}=(2,0),d=6$ supergravity coupled to just $1$ tensor multiplet and
and $n_{V}=n_{T}-1$ vector multiplets. Furthermore, the 5-dimensional theory
that one obtains by dimensional reduction of those two 6-dimensional theories
can be embedded in Heterotic String theory. 

These two 6-dimensional supergravity theories, dimensionally reduced on a
circle, are dual in the same sense in which the 10-dimensional
$\mathcal{N}=2A$ and $\mathcal{N}=2B$ supergravity theories are T-dual
\cite{Bergshoeff:1995as}, a fact related to the T-duality of the type~IIA and
IIB superstring theories compactified on circles of dual radii
\cite{Dai:1989ua,Dine:1989vu,Witten:1995ex}. Before we can interpret this
duality between supergravity theories in the context of superstring theory as
a large-small radii or coupling constant duality (for instance) we need to
find the dictionary that relates the fields of both 6-dimensional
theories. This dictionary will be the analogous of the Buscher rules for
T-duality
\cite{Buscher:1985kb,Buscher:1987sk,Buscher:1987qj,Bergshoeff:1995as,Meessen:2001wk}
and it will allow us to transform any solution of one of these theories
admitting one isometry into a solution of the dual theory. 

The initial step to derive this dictionary will be to find out how each
solution of the 5-dimensional theory can be oxidized to two different
solutions of two different 6-dimensional theories: one which only contains
chiral 2-forms and one with a non-chiral 2-form and vector fields.

To simplify the discussions, in what follows we are going to call the
6-dimensional supergravity theory with just one tensor multiplet and $n_{V}$
vector multiplets and $c_{0\ ij}=c_{1\ ij}=\delta_{ij}$, $\mathcal{N}=2A$
theories and the dual theories with $n_{T}=n_{V}+1$ tensor multiplets and no
vector multiplets, $\mathcal{N}=2B$ theories.

Now we will focus on the 5-dimensional theories with $n_{V5}=n_{V}+2$ vector
multiplets which have these two possible 6-dimensional origins.

\section{Uplifting solutions to six dimensions}
\label{sec-uplifting}

Let us consider the family of $\mathcal{N}=2,d=5$ theories coupled to
$n_{V5}=n_{V}+2$ vector multiplets and symmetric tensor $C_{IJK}$,
$I=0,\cdots, n_{V}+2$ given by $C_{0r+1s+1}=\frac{1}{3!}\eta_{r\, s}$, $r, s,
\ldots = 0,\cdots,n_{V}+1$. The scalar functions $h^{I}$ can be parametrized
in terms of the physical scalars by

\begin{equation}
\label{parametrization5}
h^{0}=2e^{-2\phi^{1}/\sqrt{6}}\, 
\hspace{1cm}
h^{r+1}=L^{r}e^{\phi^{1}/\sqrt{6}}\, ,
\end{equation}

\noindent
where the functions $L^{r}$ only depend on the scalars
$\phi^{2},\cdots,\phi^{n_{V}+2}$, and satisfy

\begin{equation}
L^{r}L^{s}\eta_{rs}=1\, .
\end{equation} 

The action can be written in terms of these functions and the scalar
$\phi^{1}$ and takes the form 

\begin{equation}
\label{5dimtheory}
\begin{aligned}
S= \int d^{5}x\sqrt{|g|}\Bigg\{
&
R+\tfrac{1}{2}(\partial\phi^{1})^{2}
-\partial_{\mu}L^{r}\partial^{\mu}L_{r}
-\tfrac{1}{48}e^{4\phi^{1}/\sqrt{6}}F^{0}F^{0}
-\tfrac{1}{6}e^{-2\phi^{1}/\sqrt{6}}\mathcal{M}_{rs}F^{r+1}F^{s+1}
\\
&
\\
&
+\frac{\epsilon}{24\sqrt{3}\sqrt{|g|}}\eta_{rs}F^{r+1}F^{s+1}A^{0}
\Bigg\}\, ,
\end{aligned}
\end{equation}

\noindent
where

\begin{equation}
L_{r}=\eta_{rs}L^{s}\, ,
\,\,\,\,
\mbox{and}
\,\,\,\,
\mathcal{M}_{rs}=2L_{r}L_{s}-\eta_{rs}\, .  
\end{equation}

For our purposes, though, it is convenient to express everything in terms of
the $h^{I}$:

\begin{equation}
L^{r}=h^{r+1}\sqrt{h^{0}/2}\, ,
\hspace{1cm}
L_{r} = h_{r+1}/\sqrt{h^{0}/2}\, ,
\hspace{1cm}
\mathcal{M}_{rs}=4\frac{h_{r+1}h_{s+1}}{h_{0}}-\eta_{rs}.
\end{equation}

According to our previous discussion, this theory can be uplifted to two
different 6-dimensional theories.

\subsection{Uplift to $\mathcal{N}=2B,d=6$ supergravity}

$\mathcal{N}=2B,d=6$ supergravity is the name that we have given to the
theories of $\mathcal{N}=(2,0),d=6$ supergravity coupled to $n_{T}=n_{V}+1$
tensor multiplets only. The equations of motion of this theory can be obtained
form the pseudoaction

\begin{equation}
\label{nTtensors}
\hat{S}=\int d^{6}\hat{x}\sqrt{|\hat{g}|}\Big\{\hat{R}
-\partial_{\hat{a}}\hat{L}^{r}\partial^{\hat{a}}\hat{L}_{r}
+\tfrac{1}{3}\hat{\mathcal{M}}_{rs}\hat{H}^{r}{}_{\hat{a} \hat{b} \hat{c}}
\hat{H}^{s\ \hat{a} \hat{b} \hat{c}}\Big\}\, ,
\end{equation}

\noindent
supplemented by the (anti-) selfduality conditions

\begin{equation}
\label{duality3}
\hat{\mathcal{M}}_{rs}\hat{H}^{r}=-\eta_{rs}\star\hat{H}^{s}\, .
\end{equation}

Then, according to the results in Section~\ref{reduction}, the 6-dimensional
fields of this theory can be expressed in terms of those of 
the 5-dimensional theory Eq.~(\ref{5dimtheory}) as follows:

\subsubsection*{Scalars}

The physical scalars $\hat{\varphi}^{\underline{\alpha}}$, and the functions
$\hat{L}^{r}$, with $\underline{\alpha}=1,\cdots,n_{V}+1$ and
$r=0,\cdots,n_{V}+1$, are given by

\begin{equation}
\label{scalarhat}
\begin{array}{rcl}
\hat{\varphi}^{\underline{\alpha}}
& = & 
\phi^{\underline{\alpha}+1}\, ,
\\
& &  \\
\hat{L}^{r}(\varphi^{\underline{\alpha}})
& = &
h^{r+1}( h^{0}/2 )^{1/2}\, . 
\end{array}
\end{equation}

\subsubsection*{Metric}

The 6-dimensional metric components are the following

\begin{equation}
  \begin{array}{rcl}
\hat{g}_{\underline{z}\underline{z}} 
& = & 
-\left(h^{0}/2\right)^{-3/2}\, ,
\\
& & \\
\hat{g}_{\mu\underline{z}} 
& = & 
-\tfrac{1}{\sqrt{12}}\left(h^{0}/2\right)^{-3/2}A^{0}{}_{\mu}\, ,
\\
& & \\
\hat{g}_{\mu\nu} 
& = & 
( h^{0}/2 )^{1/2}g_{\mu\nu}
-\tfrac{1}{12}\left(h^{0}/2\right)^{-3/2}A^{0}{}_{\mu}A^{0}{}_{\nu}\, ,
\end{array}
\end{equation}

\noindent
or, equivalently

\begin{equation}
\label{metrichat}
d\hat{s}^{2}
=
-(h^{0}/2)^{-3/2}\Big[dz+\tfrac{1}{\sqrt{12}}A^{0}\Big]^{2}
+( h^{0}/2 )^{1/2}ds^{2}\, .
\end{equation}

\subsubsection*{2-forms}

We only need to know the component $\hat{B}^{r}{}_{\mu\underline{z}}$ of the
2-forms, because the rest of components are determined through the duality
relations Eqs.~(\ref{duality3}). We have

\begin{equation}
\label{2formhat}
\hat{B}^{r}{}_{\mu\underline{z}}=\tfrac{1}{\sqrt{12}}A^{r+1}{}_{\mu}\, . 
\end{equation}

It can also be useful to have the expression of the 3-form field strengths in
the Vielbein basis:

\begin{equation}
  \begin{array}{rcl}
\hat{H}^{r}{}_{abz} 
& = & 
\tfrac{1}{\sqrt{12}}(h^{0}/2)^{2}F^{r+1}{}_{ab}\, ,
\\
& & \\
\hat{H}^{r}{}_{abc} 
& = & 
-\tfrac{1}{2\sqrt{12}}(h^{0}/2)\mathcal{M}^{r}{}_{s}
\epsilon_{abcde}F^{s+1\ de}\, ,
\end{array}
\end{equation}

\noindent
where one has to take into account that $F^{s+1\ de}$ and $\epsilon_{abcde}$
are five-dimensional quantities.

\subsection{Uplift to $\mathcal{N}=2A,d=6$ supergravity}

$\mathcal{N}=2A,d=6$ supergravity is the name that we have given to the
theories of $\mathcal{N}=(2,0),d=6$ supergravity coupled to $n_{T}=1$ tensor
multiplets and $n_{V}$ vector multiplets with $c_{0\ ij}=c_{1\
  ij}=\delta_{ij}$ with $i=1,\cdots, n_{V}$. Since in this case the two
2-forms have opposite chirality, they can be combined into a single,
unrestricted, 2-form that we are going to denote by $\tilde{B}$ (no indices)
and there is a covariant action from which one can derive directly the
equations of motion. It takes the form

\begin{equation}
\tilde{S}=\int d^{6}\tilde{x}\sqrt{|\tilde{g}|}
\left\{
\tilde{R}+\tfrac{1}{2}(\partial \tilde{\varphi})^{2}
+\tfrac{1}{3}e^{\sqrt{2}\tilde{\varphi}}\tilde{H}^{2}
-e^{\tilde{\varphi}/\sqrt{2}}\tilde{F}^{i}\tilde{F}^{i}
\right\}\, ,
\label{stringaction1}
\end{equation}

\noindent
where now we are using tildes instead of hats in order to distinguish these
fields from the previous ones and from the 5-dimensional ones. In this action,
$i=1,\cdots, n_{V}$ and the 3-form field strength is defined as

\begin{equation}
\tilde{H}=d\tilde B+\tilde{F}^{i}\wedge \tilde{A}^{i}\, .
\end{equation}

This theory is obtained when we parametrize the functions $\tilde{L}^{r}$,
$r=0,1$ as

\begin{equation}
\tilde{L}^{0}=\cosh{(\tilde{\varphi}/\sqrt{2})}\, , 
\hspace{1cm} 
\tilde{L}^{1}=\sinh{(\tilde{\varphi}/\sqrt{2})}\, ,
\end{equation}

\noindent
and $\tilde{H}$ and $\tilde{B}$ are related to the fields $\tilde{H}^{r}$ and
$\tilde{B}^{r}$ (which appear in (\ref{pseudoaction})) by

\begin{equation}
\tilde{B}=\tilde{B}^{0}-\tilde{B}^{1}\, , 
\hspace{1cm} 
\tilde{H}=\tilde{H}^{0}-\tilde{H}^{1}\, .
\end{equation}

This theory can be obtained from the compactification of $\mathcal{N}=1,d=10$
supergravity coupled to vector multiplets (the effective field theory of the
Heterotic String) on $T^{4}$ followed by a truncation. In particular, the
scalar $\tilde{\varphi}$ is related to the dilaton field of the Heterotic
String by

\begin{equation}
\tilde{\varphi} = -\sqrt{2}\, \phi_{\rm Het}\, .
\end{equation}

Now, as we have seen, this theory, also gives (\ref{5dimtheory}) when reduced
to five dimensions. In order to find the relations among the fields, we have
to use the linear transformation (\ref{lineartransform}). This gives us
directly the transformation of vector fields. Also, on taking into account the
parametrizations (\ref{parametrization5}) and (\ref{parametrization2}) we get
the relation between the different scalar fields.  This leads to the following
expressions for the 6-dimensional fields in terms of the 5-dimensional ones:

\subsubsection*{Scalar}

The dilaton is related to the five-dimensional scalars by

\begin{equation}
\label{scalartilde}
e^{\tilde{\varphi}/\sqrt{2}}=2^{-1/2}h^{0}(h^{1}+h^{2})^{1/2}\, .
\end{equation}

\subsubsection*{Metric}

The KK scalar $\phi$ and the KK vector $A_{\mu}$ are given by

\begin{equation}
e^{-2\phi/\sqrt{6}}=\tfrac{1}{2}(h^{1}+h^{2})\, ,
\hspace{1cm}
A_{\mu}=\tfrac{1}{\sqrt{12}}(A_{\mu}^{1}+A_{\mu}^{2})\, ,
\end{equation}

\noindent
and, therefore, the metric is given by

\begin{equation}
\begin{array}{rcl}
\tilde{g}_{\underline{z}\underline{z}} 
& = & 
-2^{3/2}(h^{1}+h^{2})^{-3/2}\, ,
\\
& & \\
\tilde{g}_{\mu\underline{z}} 
& = & 
-\sqrt{2/3}(h^{1}+h^{2})^{-3/2}
(A^{1}{}_{\mu}+A^{2}{}_{\mu})\, ,
\\
& & \\
\tilde{g}_{\mu\nu} & = & 
\tfrac{1}{\sqrt{2}}(h^{1}+h^{2})^{1/2}g_{\mu\nu}
-\tfrac{1}{3\sqrt{2}}(h^{1}+h^{2})^{-3/2} 
(A^{1}+A^{2})_{\mu}(A^{1}+A^{2})_{\nu}
\end{array}
\end{equation}

\noindent
or, equivalently, by

\begin{equation}
 \label{metrictilde}
d\tilde{s}^{2}
=
-2^{3/2}(h^{1}+h^{2})^{-3/2} 
\left[dz+\tfrac{1}{\sqrt{12}}(A^{1}+A^{2})\right]^{2}
+2^{-1/2}(h^{1}+h^{2})^{1/2}ds^{2}\, .
\end{equation}

\subsubsection*{Vectors}

The 1-form potentials are given by

\begin{equation}
  \begin{array}{rcl}
\tilde{A}^{i}{}_{\underline{z}} 
& = & 
{\displaystyle
-\frac{h^{i+2}}{h^{1}+h^{2}}\, ,
}
\\
& & \\
\tilde{A}^{i}{}_{\mu} 
& = & 
\tfrac{1}{\sqrt{12}}\left[A^{i+2}{}_{\mu}
+\tilde{A}^{i}{}_{\underline{z}}(A^{1}{}_{\mu}+A^{2}{}_{\mu})\right]\, ,
\end{array}
\end{equation}

\noindent
or, equivalently, by

\begin{equation}
\label{vectortilde}
\tilde{A}^{i}
= 
\tfrac{1}{\sqrt{12}}A^{i+2}-\frac{h^{i+2}}{h^{1}+h^{2}}
\left[dz+\tfrac{1}{\sqrt{12}}(A^{1}+A^{2})\right]\, . 
\end{equation}

\subsubsection*{2-form}

The components $\tilde{B}_{\mu\underline{z}}$ can be easily found to be

\begin{equation}
\label{2formtilde}
\tilde{B}_{\mu\underline{z}}
=
\tfrac{1}{\sqrt{12}}(A^{1}{}_{\mu}-A^{2}{}_{\mu})\, . 
\end{equation}

Now the components $\tilde{B}_{\mu\nu}$ are independent and have to be
explicitly given. They do not have a simple expression, though, and we must
content ourselves with the  field strength components instead:

\begin{equation}
\begin{array}{rcl}
\tilde{H}_{\mu\nu\underline{z}}
& = &
\tfrac{1}{\sqrt{3}}(h^{1}+h^{2})^{-1}
\Big\{\big[h^{1}-[h^{0}(h^{1}+h^{2})]^{-1}\big]F^{1}{}_{\mu\nu}
\\
& & \\
& & 
-\big[h^{2}+[h^{0}(h^{1}+h^{2})]^{-1}\big]F^{2}{}_{\mu\nu}
+h^{i}F^{i}{}_{\mu\nu}\Big\}\, , \qquad i\ge 3 \, ,
\\
& & \\
\tilde{H}_{\mu\nu\rho}
& = &
-\tfrac{1}{4\sqrt{3}}(h^{0})^{-2}
{\displaystyle\frac{\epsilon_{\mu\nu\rho\alpha\beta}}{\sqrt{|g|}}}
F^{0\,\alpha\beta}
+\tfrac{\sqrt{3}}{2}(A^{1}{}_{[\rho}+A^{2}{}_{[\rho})
\tilde{H}_{\mu\nu]\underline{z}}\, .
\end{array}
\end{equation}

\subsection{Uplift to $\mathcal{N}=2A^{*},d=6$ supergravity}

The theory that we have called $\mathcal{N}=2A,d=6$ supergravity is not
uniquely defined. One can obtain another theory that we are going to call
$\mathcal{N}=2A^{*},d=6$ supergravity by dualizing the field strength
$\tilde{H}$ into another field strength $\breve{H}$ given by\footnote{In the
  Einstein frame this is the only field which is modified in this
  transformation. We will denote all the field of this theory with $\breve{~}$
  accents anyway.}

\begin{equation}
\label{eq:3formduality}
  \breve{H}=-e^{\sqrt{2}{\tilde{\varphi}}}\star\tilde{H}\, .
\end{equation}

It turns out that this new field strength is an exact 3-form:

\begin{equation}
\breve{H}=d\breve{B}\, ,
\end{equation}

\noindent
and $\breve{H}$ and $\breve{B}$ are related to $\hat{H}^{r}$ and $\hat{B}^{r}$
in the theory of Eq.~(\ref{pseudoaction}) with $n_{T}=1$, arbitrary $n_{V}$ and
$c_{0\, ij}=c_{1\, ij}=\delta_{ij}$ by\footnote{Observe that the absence of
  Chern-Simons term in $\breve{H}$ is due to the cancellation of those in
  $\hat{H}^{0}$ and $\hat{H}^{1}$ and not to the vanishing of the constants
  $c^{r}{}_{ij}$.}

\begin{equation}
\breve{H}=\hat{H}^{0}+\hat{H}^{1}\, ,
\hspace{1cm}
 \breve{B}=\hat{B}^{0}+\hat{B}^{1}\, .
\end{equation}

The action for this theory is

\begin{equation}
\label{stringaction2}
\breve{S}
=
\int d^{6}\breve{x}\sqrt{|\breve{g}|}
\Bigg\{\breve{R}+\tfrac{1}{2}(\partial \breve{\varphi})^{2}
+\tfrac{1}{3}e^{-\sqrt{2}\breve{\varphi}}\breve{H}^{2}
-e^{\breve{\varphi}/\sqrt{2}}\breve{F}^{i}\breve{F}^{i}
-\frac{\epsilon}{3\sqrt{|g|}}\breve{H}\breve{F}^{i}\breve{A}^{i}
\Bigg\}\, .
\end{equation}

This theory can be obtained from the effective field theory of the type~IIA
superstrings compactified on K3
\cite{Seiberg:1988pf,Duff:1993ij,Hull:1994ys,Duff:1994zt,Witten:1995ex}
followed by a truncation. In particular,  the scalar $\breve{\varphi}$ (which
is equal to $\tilde{\varphi}$), is related to the dilaton of that superstring
theory by

\begin{equation}
\breve{\varphi} = \sqrt{2}\, \phi_{IIA}\, .  
\end{equation}

The different coupling of the dilaton field to the vector fields (comparing
with the $\mathcal{N}=2A$ case) is mainly due to the fact that they are RR
fields in this case instead of  NSNS fields.

All the fields have the same relation with the five-dimensional ones as the
tilded ones, except for the 2-form $\breve{B}$, whose components
$\mu\underline{z}$ now are given by

\begin{equation}
\breve{B}_{\mu\underline{z}}=\tfrac{1}{\sqrt{12}}A^{0}{}_{\mu}\, .
\end{equation}

The 3-form field strength is given by

\begin{equation}
\begin{array}{rcl}
\breve{H}_{\mu\nu\underline{z}} 
& = & 
\tfrac{1}{\sqrt{12}} F^{0}{}_{\mu\nu}\, ,
\\
& & \\
\breve{H}_{\mu\nu\rho}
& = & 
-\tfrac{1}{8\sqrt{3}}
(h^{0})^{2}(h^{1}+h^{2})
{\displaystyle\frac{\epsilon_{\mu\nu\rho\alpha\beta}}{\sqrt{|g|}}}
\Big\{\big[h^{1}-[h^{0}(h^{1}+h^{2})]^{-1}\big]F^{1\, \alpha\beta}\\
& & \\
& & 
-\big[h^2+[h^0(h^1+h^2)]^{-1}\big]F^{2\, \alpha\beta}+h^{i}F^{i\, \alpha\beta}\Big\}
+\tfrac{\sqrt{3}}{2}(A^{1}{}_{[\rho}+A^{2}{}_{[\rho})
\breve{H}_{\mu\nu]\underline{z}}\, , \qquad i\ge 3\, .
\end{array}
\end{equation}

\section{Maps between six-dimensional theories}
\label{sec-maps}

Putting together all our results we can write the following generalization of
the Buscher rules between the $\mathcal{N}=2A,2A^{*}$ and $2B$ theories:

\subsection*{From $\mathcal{N}=2B$ to $\mathcal{N}=2A$}

\begin{equation}
  \begin{array}{rcl}
e^{\sqrt{2}\tilde{\varphi}} 
& = & 
-2(\hat{L}^{0}+\hat{L}^{1})/\hat{g}_{\underline{z}\underline{z}}\, ,
\\
& & \\
\tilde{g}_{\underline{z}\underline{z}} 
& = & 
-2^{3/2}(\hat{L}^{0}+\hat{L}^{1})^{-3/2}
|\hat{g}_{\underline{z}\underline{z}}|^{-1/2}\, ,
\\
& & \\
\tilde{g}_{\mu\underline{z}} 
& = & 
-2^{3/2}(\hat{L}^{0}+\hat{L}^{1})^{-3/2}|\hat{g}_{\underline{z}\underline{z}}|^{-1/2}
(\hat{B}^{0}+\hat{B}^{1})_{\mu\underline{z}}\, ,
\\
& & \\
\tilde{g}_{\mu\nu} 
& = & 
2^{-1/2}(\hat{L}^{0}+\hat{L}^{1})^{1/2}
\Big[|\hat{g}_{\underline{z}\underline{z}}|^{1/2}\hat{g}_{\mu\nu}
+|\hat{g}_{\underline{z}\underline{z}}|^{-1/2}\hat{g}_{\mu\underline{z}}
\hat{g}_{\nu\underline{z}}\Big] \\
& & \\
& & 
-2^{3/2}(\hat{L}^{0}+\hat{L}^{1})^{-3/2}|\hat{g}_{\underline{z}\underline{z}}|^{-1/2}
(\hat{B}^{0}+\hat{B}^{1})_{\mu\underline{z}}
(\hat{B}^{0}+\hat{B}^{1})_{\nu\underline{z}}\, ,
\\
& & \\
\tilde{A}^{i}{}_{\underline{z}} 
& = & 
-\hat{L}^{i+1}/(\hat{L}^{0}+\hat{L}^{1})\, ,
\\
& & \\
\tilde{A}^{i}{}_{\mu} 
& = &
\hat{B}^{i+1}{}_{\mu\underline{z}}
-\hat{L}^{i+1}(\hat{B}^{0}+\hat{B}^{1})_{\mu\underline{z}}/(\hat{L}^{0}+\hat{L}^{1})\, ,
\\
& & \\
\tilde{B}_{\mu \underline{z}}  
& = &  
(\hat{B}^{0}-\hat{B}^{1})_{\mu \underline{z}}\, .
\end{array}
\end{equation}

\subsection*{From $\mathcal{N}=2A$ to $\mathcal{N}=2B$}

\begin{equation}
  \begin{array}{rcl}
\left| \hat{g}_{\underline{z} \underline{z}} \right| 
& = & 
2^{\frac{3}{2}} e^{-\frac{3}{2 \sqrt{2}} \tilde{\varphi}} 
\left|\tilde{g}_{\underline{z} \underline{z}}\right|^{-\frac{1}{2}} \, ,
\\
& & \\
\hat{g}_{\mu \underline{z}} 
& = & 
-2^{\frac{3}{2}} e^{-\frac{3}{2 \sqrt{2}} \tilde{\varphi}} 
\left|\tilde{g}_{\underline{z} \underline{z}}\right|^{-\frac{1}{2}} 
(\tilde{B}^{0}+\tilde{B}^{1})_{\mu \underline{z}} \, ,
\\
& & \\
\hat{g}_{\mu \nu} 
& = & 
2^{-\frac{1}{2}} \left| \tilde{g}_{\underline{z}
    \underline{z}}\right|^{\frac{1}{2}} 
e^{\frac{\tilde{\varphi}}{2 \sqrt{2}}} 
\left( \tilde{g}_{\mu \nu}-\tilde{g}_{\mu \underline{z}} 
\tilde{g}_{\nu \underline{z}}/\tilde{g}_{\underline{z} \underline{z}}
\right)
+2^{\frac{3}{2}} e^{- \frac{3}{2 \sqrt{2}} \tilde{\varphi}} 
\left|\tilde{g}_{\underline{z} \underline{z}}\right|^{-\frac{1}{2}} 
(\tilde{B}^{0}+\tilde{B}^{1})_{\mu \underline{z}} 
(\tilde{B}^{0}+\tilde{B}^{1})_{\nu \underline{z}} \, , 
\\
& & \\
\hat{L}^{0}  
& = & 
2^{-\frac{3}{2}} e^{-\frac{\tilde{\varphi}}{2 \sqrt{2}}} 
\left| \tilde{g}_{\underline{z} \underline{z}} \right|^{\frac{1}{2}} 
+ 2^{-\frac{1}{2}} e^{\frac{\tilde{\varphi}}{2 \sqrt{2}}} 
\left| \tilde{g}_{\underline{z} \underline{z}} \right|^{-\frac{1}{2}} 
\left(1+\tilde{A}^{r}{}_{\underline{z}} \tilde{A}^{r}{}_{\underline{z}}\right)\, ,
\qquad r>1 \, ,
\\
& & \\
\hat{L}^{1} 
& = &
-2^{-\frac{3}{2}} e^{-\frac{\tilde{\varphi}}{2 \sqrt{2}}} 
\left| \tilde{g}_{\underline{z} \underline{z}} \right|^{\frac{1}{2}} 
+ 2^{-\frac{1}{2}} e^{\frac{\tilde{\varphi}}{2 \sqrt{2}}} 
\left| \tilde{g}_{\underline{z} \underline{z}} \right|^{-\frac{1}{2}} 
\left(1-\tilde{A}^{r}{}_{\underline{z}} \tilde{A}^{r}{}_{\underline{z}}\right)\, ,
\qquad r>1 \, ,
\\
& & \\
\hat{L}^{r} 
& = & 
-\sqrt{2} \left| \tilde{g}_{\underline{z} \underline{z}}\right|^{-\frac{1}{2}} 
e^{\frac{\tilde{\varphi}}{2 \sqrt{2}}} \tilde{A}^{r-1}{}_{\underline{z}}\, , 
\qquad r\ge 2 \, , 
\\
& & \\
\hat{B}^{0}{}_{\mu \underline{z}} 
& = & 
\tfrac{1}{2} \left(\tilde{B}_{\mu \underline{z}} 
+\tilde{g}_{\mu \underline{z}}/\tilde{g}_{\underline{z} \underline{z}}\right)
\, , 
\\
& & \\
\hat{B}^{1}{}_{\mu \underline{z}} 
& = & 
\tfrac{1}{2} \left(-\tilde{B}_{\mu \underline{z}} 
+\tilde{g}_{\mu \underline{z}}/\tilde{g}_{\underline{z} \underline{z}}\right)
\, , 
\\
& & \\  
\hat{B}^{r}{}_{\mu \underline{z}} 
& = & \tilde{A}^{r-1}{}_{\mu} - \tilde{A}^{r-1}{}_{\underline{z}}\, 
\tilde{g}_{\mu \underline{z}}/\tilde{g}_{\underline{z} \underline{z}}\, , 
\qquad r\ge 2 \, .
\end{array}
\end{equation}  

\subsection{From $\mathcal{N}=2B$ to $\mathcal{N}=2A^{*}$}

\begin{equation}
  \begin{array}{rcl}
e^{\sqrt{2}\breve{\varphi}} 
& = & 
-2(\hat{L}^{0}+\hat{L}^{1})/\hat{g}_{\underline{z}\underline{z}}\, ,
\\
& & \\
\breve{g}_{\underline{z}\underline{z}} 
& = & 
-2^{3/2}(\hat{L}^{0}+\hat{L}^{1})^{-3/2}
|\hat{g}_{\underline{z}\underline{z}}|^{-1/2}\, ,
\\
& & \\
\breve{g}_{\mu\underline{z}} 
& = & 
-2^{3/2}(\hat{L}^{0}+\hat{L}^{1})^{-3/2}|\hat{g}_{\underline{z}\underline{z}}|^{-1/2}
(\hat{B}^{0}+\hat{B}^{1})_{\mu\underline{z}}\, ,
\\
& & \\
\breve{g}_{\mu\nu} 
& = & 
2^{-1/2}(\hat{L}^{0}+\hat{L}^{1})^{1/2}
\Big[|\hat{g}_{\underline{z}\underline{z}}|^{1/2}\hat{g}_{\mu\nu}
+|\hat{g}_{\underline{z}\underline{z}}|^{-1/2}
\hat{g}_{\mu\underline{z}}\hat{g}_{\nu\underline{z}}\Big]
\\
& & \\
& &
-2^{3/2}(\hat{L}^{0}+\hat{L}^{1})^{-3/2}|\hat{g}_{\underline{z}\underline{z}}|^{-1/2}
(\hat{B}^{0}+\hat{B}^{1})_{\mu\underline{z}}
(\hat{B}^{0}+\hat{B}^{1})_{\nu\underline{z}}\, ,
\\
& & \\
\breve{A}^{i}{}_{\underline{z}} 
& = & 
-\hat{L}^{i+1}/(\hat{L}^{0}+\hat{L}^{1})\, ,
\\
& & \\
\breve{A}^{i}{}_{\mu} 
& = & 
\hat{B}^{i+1}{}_{\mu\underline{z}}
-\hat{L}^{i+1}(\hat{B}^{0}+\hat{B}^{1})_{\mu\underline{z}}/
(\hat{L}^{0}+\hat{L}^{1})\, ,
\\
& & \\
\breve{B}_{\mu\underline{z}} 
& = & 
\hat{g}_{\mu\underline{z}}/\hat{g}_{\underline{z}\underline{z}}\, .
\end{array}
\end{equation}

\subsection*{From $\mathcal{N}=2A^{*}$ to $\mathcal{N}=2B$}

\begin{equation}
\begin{array}{rcl}
\left| \hat{g}_{\underline{z} \underline{z}} \right|  
& = &  
2^{\frac{3}{2}} e^{-\frac{3}{2 \sqrt{2}} \breve{\varphi}}
\left|\breve{g}_{\underline{z} \underline{z}}\right|^{-\frac{1}{2}} \, ,
\\
& & \\
\hat{g}_{\mu \underline{z}}  
& = & 
-2^{\frac{3}{2}} e^{-\frac{3}{2 \sqrt{2}} \breve{\varphi}}
\left|\breve{g}_{\underline{z} \underline{z}}\right|^{-\frac{1}{2}}
\breve{B}_{\mu \underline{z}} \, ,
\\
& & \\
\hat{g}_{\mu \nu}  
& = &  
2^{-\frac{1}{2}} \left| \breve{g}_{\underline{z}\underline{z}}\right|^{\frac{1}{2}} 
e^{\frac{\breve{\varphi}}{2 \sqrt{2}}} 
\left( 
\breve{g}_{\mu\nu}
-\breve{g}_{\mu\underline{z}}\breve{g}_{\nu\underline{z}}/\breve{g}_{\underline{z}\underline{z}} 
\right)
+2^{\frac{3}{2}} e^{-\frac{3}{2 \sqrt{2}} \breve{\varphi}} 
\left|\breve{g}_{\underline{z} \underline{z}}\right|^{-\frac{1}{2}} 
\breve{B}_{\mu \underline{z}} \breve{B}_{\nu \underline{z}} \, , 
\\
& & \\
\hat{L}^{0}   
& = &  
2^{-\frac{3}{2}} e^{-\frac{\breve{\varphi}}{2 \sqrt{2}}} 
\left| \breve{g}_{\underline{z} \underline{z}} \right|^{\frac{1}{2}} 
+2^{-\frac{1}{2}} e^{\frac{\breve{\varphi}}{2 \sqrt{2}}} 
\left| \breve{g}_{\underline{z} \underline{z}} \right|^{-\frac{1}{2}} 
\left(1+\breve{A}^{r}{}_{\underline{z}}
  \breve{A}^{r}{}_{\underline{z}}\right)\, , \qquad r>1 \, ,
\\
& & \\
\hat{L}^{1}  
& = & 
-2^{-\frac{3}{2}} e^{-\frac{\breve{\varphi}}{2 \sqrt{2}}} 
\left| \breve{g}_{\underline{z} \underline{z}} \right|^{\frac{1}{2}} 
+2^{-\frac{1}{2}} e^{\frac{\breve{\varphi}}{2 \sqrt{2}}} 
\left| \breve{g}_{\underline{z} \underline{z}} \right|^{-\frac{1}{2}} 
\left(1-\breve{A}^{r}{}_{\underline{z}} \breve{A}^{r}{}_{\underline{z}}\right)\, ,
\qquad r>1 \, ,
\\
& & \\  
\hat{L}^{r}  
& = &  
-\sqrt{2} \left| \breve{g}_{\underline{z} \underline{z}}\right|^{-\frac{1}{2}} 
e^{\frac{\breve{\varphi}}{2 \sqrt{2}}} \breve{A}^{r-1}_{\underline{z}} \, ,
\qquad r \ge 2 \, , 
\\
& & \\
\hat{B}^{0}{}_{\mu \underline{z}}  
& = &  
\tfrac{1}{2} \left(\tilde{B}_{\mu \underline{z}} 
+\breve{g}_{\mu \underline{z}}/\breve{g}_{\underline{z} \underline{z}}\right)\, , 
\\
& & \\
\hat{B}^{1}{}_{\mu \underline{z}}  
& = &  
\tfrac{1}{2} 
\left(-\tilde{B}_{\mu \underline{z}} 
+\breve{g}_{\mu \underline{z}}/\breve{g}_{\underline{z} \underline{z}}\right)
\, , 
\\
& & \\
\hat{B}^{r}{}_{\mu \underline{z}}  
& = &  
\breve{A}^{r-1}{}_{\mu} - \breve{A}^{r-1}{}_{\underline{z}} 
\breve{g}_{\mu \underline{z}}/\breve{g}_{\underline{z} \underline{z}}\, ,
\qquad r\ge 2 \, .
 \end{array}
\end{equation}

\section{Applications}
\label{sec-applications}

We are now ready to exploit the relations between 5- and 6-dimensional
theories that we have uncovered. There is one more twist that we can add to
them, though: observe that if we had dimensionally reduced the \textit{gauged}
$\mathcal{N}=2A,d=6$ theory we would have obtained a gauged
$\mathcal{N}=2,d=5$ supergravity theory and the relation between the physical
fields of these two gauged theories would be exactly the same we have obtained
in the ungauged case. This is true as long as the gauge group does not change
in the process of dimensional reduction (as in the case of generalized
dimensional reduction \cite{Scherk:1979zr}). Then, we can use the formulae we
have obtained to uplift solutions of the 5-dimensional gauged theories to
solutions of the 6-dimensional gauged theories and vice-versa.

There are some points to be discussed and clarified before carrying out this
program. 

First of all we must discuss the possible gaugings of these theories.  
The $\mathcal{N}=2A,d=6$ theories can be gauged in essentially two ways:

\begin{enumerate}

\item We could just gauge a subgroup of the $\mathrm{SO}(n_{V})$ group that
  rotates the vector fields among themselves. The only fermion fields this
  global symmetry acts on are the gaugini, which carry the same indices as the
  vector fields and an $\mathrm{Sp}(1)\sim \mathrm{SU}(2)$ R-symmetry index
  which remains inert under these transformations. Observe that the only
  scalar of the theory, the dilaton, is also inert.

\item We can gauge the whole R-symmetry group, $\mathrm{SO}(3)$ or a
  $\mathrm{SO}(2)$ subgroup of it using Fayet-Iliopoulos terms.\footnote{This
    is the theory considered in Ref.~\cite{Cariglia:2004kk}, for instance. }
  Observe that one needs vectors transforming in the same fashion. Thus, in
  this case one would be gauging $\mathrm{SO}(3)$ or a $\mathrm{SO}(2)$
  subgroup of $\mathrm{SO}(n_{V})$ which, on top of acting on some the
  $\mathrm{SO}(n_{V})$ indices of the vectors and gaugini, would also act on
  the R-symmetry indices of all the fermions of the theory, which would now be
  charged.

\end{enumerate} 

The dimensional reduction of these gauged 6-dimensional theories would be the
models of $\mathcal{N}=1,d=5$ supergravity that we have found, characterized
by the $C_{IJK}$ tensor with non-vanishing indices
$C_{0rs}=\tfrac{1}{3!}\eta_{rs}$, with exactly the same kind of gaugings (with
or without Fayet-Iliopoulos terms). The main difference with the
6-dimensional theories is that, in the non-Abelian case, the gauge group acts
on the scalars that originate in the 6th component of the 6-dimensional vector
fields and these transformations are isometries of the $\sigma$-model metric.
The relations between 5- and 6-dimensional fields can be used directly in the
gauged case but we must take into account that in order to get the $C_{IJK}$
tensor in the form $C_{0rs}=\tfrac{1}{3!}\eta_{rs}$ we had to make linear
combinations of several different vector fields. This can only be done if they
have the same transformation properties under the group to be gauged, which is
not the case. Thus, we only must gauge vector fields not involved in these
redefinitions.

The $\mathcal{N}=2B,d=6$ theories cannot be gauged, at least in a conventional
way. However, it is believed that there are 6-dimensional gauge theories based
on chiral 2-forms associated to coincident M5-branes. The main reason is that,
when compactified on a circle, M5-branes behave as D4-branes and the
Born-Infeld fields of coincident D4-branes are non-Abelian. This means that,
at least, the non-Abelian theory of 2-forms exists when one of the 6
dimensions is compactified on a circle and, in those conditions, the massless
modes are essentially non-Abelian 1-forms. Actually, there have been several
proposals of non-Abelian theories of 2-forms in 6 dimensions
\cite{Hofman:2002ey,Ho:2011ni,Huang:2012tu} and, in general, they consider
that one of the 6 dimensions is compactified.

The situation we are facing here is similar and, probably, directly related to
the worldvolume theories of the M5-branes.  It is clear that, when these
theories are compactified on a circle, at least the massless part of the
spectrum (1-forms in $d=5$) can be gauged. We do not know how to formulate the
gauging using chiral 2-forms directly in 6 uncompactified dimensions but we do
known that, at lowest order, the relation between the 6- and 5-dimensional
non=Abelian fields is the same as between the Abelian ones. We can, therefore,
use the uplifting formulae to construct non-Abelian solutions of a
``$\mathrm{SO}(3)$-gauged'' $\mathcal{N}=2B,d=6$ theory whose exact
6-dimensional formulation we do not know. Actually, we can use this relation
as a lowest-order formulation of that theory which probably only exists anyway
when one of the 6 dimensions is compactified on a circle.

\subsection{Solutions of the  $\mathrm{SO}(3)$-gauged $\mathcal{N}=2A^{*},d=6$ theory}

The supersymmetric solutions of the gauged $\mathcal{N}=2A,d=6$ theory with
Fayet-Iliopoulos (FI) terms were classified in Ref.~\cite{Cariglia:2004kk},
where some interesting examples were also constructed. We can dimensionally
reduced them to 5 dimensions using our results but we prefer to construct
supersymmetric solutions of the $\mathrm{SO}(3)$-gauged $\mathcal{N}=2A,d=6$
theory without FI terms by uplifting some of the supersymmetric solutions of
the similarly gauged (no FI terms) $\mathcal{N}=2,d=5$ supergravity with no
hypermultiplets\footnote{These theories are the simplest supersymmetrization
  of the Einstein-Yang-Mills (EYM) theory and have been called
  $\mathcal{N}=2,d=5$ Super-Einstein-Yang-Mills (SEYM) theories in
  Ref. ~\cite{Meessen:2015enl}. They are related by dimensional reduction to
  the $\mathcal{N}=2,d=4$ SEYM theories
  \cite{Huebscher:2007hj,Meessen:2008kb,Hubscher:2008yz,Bueno:2014mea,Bueno:2015wva,Meessen:2015nla}. The
  same relation applies to the 4- and 5-dimensional solutions.}  recently
constructed in Ref.~\cite{Meessen:2015enl}. In particular, we are going to
uplift an extremal black hole sourced by a BPST instanton
\cite{Belavin:1975fg}.

Thus, let us consider the $\mathcal{N}=2,d=5$ SEYM theory with $n_{V5}=5$
vectors labeled by $x=1,\cdots,5$ or $x=1,2,A$ where $A,B,\ldots$ label the
three directions gauged with the group $\mathrm{SO}(3)$ and with non-vanishing
components of $C_{IJK}$ given by $C_{0xy}=\tfrac{1}{3!}\eta_{xy}$,
$\eta=\mathrm{diag}(+-----)$. The solution that we are going to uplift was
obtained in a model with one vector multiplet less but, here, for the reasons
explained above, we cannot gauge the first vector multiplets and so we add one
more ($x=2$) whose fields will vanish identically.

The metric is static and spherically symmetric

\begin{equation}
d s^{2} = f^{2} d t^{2}- f^{-1} ( d \rho^{2} + \rho^{2} d \Omega_{(3)}^{2} )\, ,
\end{equation}

\noindent
where the metric function $f$ is given by 

\begin{equation}
f^{-1} 
=
3\cdot 2^{-1/3} 
\left\{ L_{1}^{2} \left[ L_{0} - \frac{9}{2g^{2}}
    \left(\rho+\frac{\lambda^{2}}{4} \rho^{3}\right)^{-2} \right] 
\right\}^{1/3} \, ,  
\end{equation}

\noindent
where $L_{0}$ and $L_{1}$ are two spherically symmetric harmonic
functions\footnote{Not to be confused with the 6-dimensional scalar functions
  $\hat{L}_{r}$.} on $\mathbb{R}^{4}$

\begin{equation}
 L_{0,1}=a_{0,1} +q_{0,1}/\rho^{2}\, ,
\end{equation}

\noindent
$a_{0,1}$ being integration constants and $q_{0,1}$ being electric
charges. The integration constants are constrained by the normalization of the
metric at infinity, but we are are not going to impose this condition in 5
dimensions. 

There is only one non-trivial scalar that we can write as $h^{1}/h^{0}$, for
instance. In terms of the scalar functions $h^{I}$ we have

\begin{eqnarray}
 h^{0} 
& = &
2^{-1/3}\
\left[ 
\frac{L_{1}}{L_{0}- \frac{9}{2g^{2}} \, 
\left(\rho+\frac{\lambda^{2}}{4} \rho^{3}\right)^{-2}} 
\right]^{2/3}\, ,
\\
& & \nonumber \\
h^{1} 
& = &
2^{2/3}
\left[ 
\frac{L_{1}}{L_{0}- \frac{9}{2g^{2}} \, 
\left(\rho+\frac{\lambda^{2}}{4} \rho^{3}\right)^{-2}} 
\right]^{-1/3}\, ,
\\
& & \nonumber \\
h^{2} & = & h^{A} = 0 \, ,
\end{eqnarray}

\noindent
and 

\begin{equation}
\phi^{1}
=
2\frac{L_{0}- \frac{9}{2g^{2}} \, 
\left(\rho+\frac{\lambda^{2}}{4} \rho^{3}\right)^{-2}}{L_{1}}\, .
\end{equation}

Finally, the vector fields of the solution are given by

\begin{equation}
\begin{array}{rcl}
A^{0} 
& = &  
-\tfrac{1}{\sqrt{3}}
\left[L_{0}- \frac{9}{2g^{2}} \, 
\left(\rho+\frac{\lambda^{2}}{4} \rho^{3}\right)^{-2}\right]^{-1}dt \, ,
\\
& & \\
A^{1} 
& = &  
-\tfrac{2}{\sqrt{3}}  L_{1}^{-1} \, d t \, ,
\\
& & \\
A^{2} 
& = &  0 \, ,
\\
& & \\
A^{A} 
& = &  
-\frac{1}{g}\, \left(1+\frac{\lambda^{2}}{4} \rho^{2}\right)^{-1} \, v^{A}{}_{L}\, ,
\end{array}
\end{equation}

\noindent
where the $v^{A}{}_{L}$ are the left-invariant Maurer-Cartan 1-forms of the
Lie group $\mathrm{SU}(2)$, given in our conventions in the Appendix of
Ref.~\cite{Bueno:2015wva}. $A^{A}$ is the potential of the BPST instanton and
$g$ is the 5-dimensional gauge coupling constant.

It is now straightforward to uplift this solution to a solution of the
$\mathcal{N}=2A,d=6$ theory with $n_{T}=1$ (by definition) and
$n_{V}=n_{V5}-2=3$ (one of the six 5-dimensional vectors is the KK vector and
the other two come from the non-chiral 2-form) and the 3 vectors are the gauge
field of the $\mathrm{SO}(3)$ gauge group\footnote{Globally, the instanton
  solution requires the group to be $\mathrm{SU}(2)$.}

Using Eqs.~(\ref{scalartilde}),(\ref{metrictilde}),(\ref{vectortilde}) and
(\ref{2formtilde}), we find the following 6-dimensional fields:\footnote{We
  have renamed the coordinates $z$ and $t$ as $u$ and $v$, respectively, since
  they are conjugate null coordinates in 6 dimensions. Then, we have shifted
  one of them $v=v^{\prime}+\tfrac{3}{2}a_{1}u$. The null coordinates $u$ and
  $v^{\prime}$ can be expressed in terms of time ($\tau$) and space ($y$)
  coordinate as
  \begin{equation}
  u= \tfrac{1}{\sqrt{2}}(\tau +y)\, , 
  \hspace{1cm}  
  v= \tfrac{1}{\sqrt{2}}(\tau -y)\, .
  \end{equation}
}

\begin{equation}
  \begin{array}{rcl}
d\breve{s}^{2} 
& = & 
2\breve{f} du\left[dv^{\prime}-\tfrac{3}{2}(L_{1}-a_{1}) du\right]
-\breve{f}^{-1}(d\rho^{2}+\rho^{2} d\Omega_{(3)}^{2})\, ,
\\
& & \\
\breve{f}
& = &
\tfrac{\sqrt{2}}{3}
\left\{
L_{1}
\left[
L_{0}- \frac{2}{9g^{2}} \, 
\left(\rho+\frac{\lambda^{2}}{4} \rho^{3}\right)^{-2}\right]
\right\}^{-1/2}\, ,
\\
& & \\
e^{\sqrt{2} \breve{\varphi}}  
& = &  
\tfrac{1}{2}L_{1}\left[L_{0}- \frac{2}{9 g^{2}} \, 
\left(\rho+\frac{\lambda^{2}}{4} \rho^{3}\right)^{-2}\right]^{-1}\, ,
\\
& & \\
\breve{A}^{A}  
& = &  
-\frac{1}{\sqrt{12} g} \, 
\left(1+\frac{\lambda^{2}}{4} \rho^{2}\right)^{-1} \, v^{A}{}_{L} \, , 
\\
& & \\
\breve{H} 
& = & 
-\tfrac{1}{6} dv^{\prime}\wedge du \wedge
d
\left[L_{0}- \frac{2}{9g^{2}} \, 
\left(\rho+\frac{\lambda^{2}}{4} \rho^{3}\right)^{-2}\right]^{-1}
+\tfrac{3}{2}q_{1} \omega_{3}\, ,
\end{array}
\end{equation}

\noindent
and where $\omega_{3}$ is the volume form of the round 3-sphere of unit radius
whose metric is $d\Omega_{(3)}^{2}$. If, for instance, we use the Euler
coordinates $(\theta,\phi,\psi)$ such that

\begin{equation}
d\Omega_{(3)}^{2}= 
\tfrac{1}{4}
\left[
(d\psi +\cos{\theta}d\phi)^{2}+ d\theta^{2}+\sin^{2}{\theta}d\phi^{2}
\right]\, ,
\end{equation}

\noindent
then $\omega_{3}=\tfrac{1}{8}\sin{\theta} d\theta\wedge d\phi\wedge d\psi$,
and the 2-form $\breve{B}$ can be written in this coordinate patch,
up to gauge transformations, as

\begin{equation}
\label{eq:BreveB}
\breve{B}
=
-\tfrac{1}{6}
\left[L_{0}- \tfrac{2}{9g^{2}} \, 
\left(\rho+\frac{\lambda^{2}}{4} \rho^{3}\right)^{-2}\right]^{-1}
dv^{\prime}\wedge du
+\tfrac{3}{16}q_{1}\cos{\theta} d\psi\wedge
d\phi\, .
\end{equation}

Observe  that now $\breve{A}^{A}$ carries a factor of $1/\sqrt{12}$ with
respect to the potential of the BPST instanton. The reason behind this
apparent inconsistency is that the rescaling of the potentials is harmless in
the Abelian case but brings the non-Abelian 2-form field strength to an
unconventional form. To bring it back to the standard form we just have to
rescale the coupling constant. Thus, the 6-dimensional coupling constant is
given in terms of the 5-dimensional one by

\begin{equation}
\breve{g} = \sqrt{12}g\, .  
\end{equation}

The metric $d\breve{s}^{2}$ is typical that of a superposition of a string
lying in the $z$ direction and a wave with momentum $\sim q_{1}$ in the same
direction. The 3-form field strength $\breve{H}$ indicates that the string is
dyonic, with electric and magnetic charges $\sim q_{0}, q_{1}$. This kind of
solutions are very well known as they are particular cases of 3-charge
configurations dual to the D1D6W one.\footnote{Only two out of the three
  different charges are independent in this solution. This is necessary to
  have a consistent truncation to minimal supergravity.} The additional
ingredient here is the BPST instanton that modifies the metric function
$\breve{f}$. The string part of this solution is also clearly related to the
``gauge dyonic string'' solution of the Heterotic string effective action
compactified to 6 dimensions constructed in Ref.~\cite{Duff:1996cf} by adding
Yang-Mills instantons in the transverse directions to the dyonic string found
in Ref.~\cite{Duff:1995yh} (see also Ref.~\cite{Duff:1997me}).

We have left intentionally undetermined the integration constants $a_{0},a_{1}$
because different choices can leave, as we are going to see, to physically
inequivalent solutions, depending on whether we demand asymptotic flatness or
not.

\subsubsection*{Asymptotic limit}

Let us first consider the $\rho\rightarrow\infty$ limit. There are two
possibilities:

\begin{enumerate}
\item If we choose the two integration constants in the harmonic functions
  $L_{0,1}$ to be non-vanishing, $a_{0}a_{1}> 0$

\begin{equation}
\breve{f}\sim \frac{\sqrt{2}/3}{\sqrt{a_{0}a_{1}}}
\, ,
\hspace{.7cm}
e^{\sqrt{2}\breve{\varphi}_{\infty}}= \frac{a_{1}}{2a_{0}}\, ,
\,\,\,\,
\mbox{and}
\,\,\,\,
\breve{H}_{\rho v^{\prime} u} \sim -\frac{q_{0}}{3a_{0}^{2}}
\frac{1}{\rho^{3}}\, .
\end{equation}

\noindent
First of all, we see that the metric is asymptotically flat. The normalization
$\breve{f}=1$ fixes the integration constants in terms of just
$\breve{\varphi}_{\infty}$:

\begin{equation}
a_{0}= \tfrac{1}{3}e^{-\breve{\varphi}{\sqrt{2}}}\, ,
\hspace{1cm}  
a_{1}= \tfrac{2}{3}e^{+\breve{\varphi}{\sqrt{2}}}\, .
\end{equation}

This solution describes the superposition of the dyonic string and $pp$-wave
mentioned above. The charges of the string can be easily computed and are
given by

\begin{equation}
Q
\equiv 
\tfrac{1}{2\pi^{2}}\int_{S^{3}_{\infty}}e^{-\sqrt{2}\breve{\varphi}}\star \breve{H}
= 
-3q_{0}\, ,
\hspace{1cm}
P 
\equiv  
\tfrac{1}{2\pi^{2}}\int_{S^{3}_{\infty}} \breve{H} =\tfrac{3}{2} q_{1}\, .
\end{equation}

The instanton field falls too fast at infinity to give any contributions to
charges, masses or momenta.

\item If both integration constants vanish $a_{0}=a_{1}=0$\footnote{If only
    one of them vanished, the dilaton would not be well defined.}, as long as
  $q_{1}\left(q_{0}-\frac{8}{3 \breve{g}^{2}}\right)$ $\breve{f}$ remains
  always finite and strictly real and positive for all finite values of $\rho$
  and the whole metric is regular. In the $\rho \rightarrow \infty$ limit the
  fields behave as

\begin{equation}
\breve{f}
\sim
 \frac{\rho^{2}}{R^{2}_{\infty}}\, ,
\hspace{.7cm}
e^{\sqrt{2}\breve{\varphi}_{\infty}}=\frac{q_{1}}{2q_{0}}\, ,
\,\,\,\,
\mbox{and}
\,\,\,\,
\breve{H}_{\rho v^{\prime} u} 
\sim
-\frac{1}{3q_{0}} \rho\, ,
\end{equation}

\noindent
where we have defined the constant

\begin{equation}
\label{eq:Rinfty}
R^{2}_{\infty}\equiv \sqrt{\frac{9q_{0}q_{1}}{2}}\, ,
\end{equation}

\noindent
which depends on the charges but not on the modulus
$\breve{\varphi}_{\infty}$, and the metric takes a direct product form

\begin{equation}
d\breve{s}_{\infty}^{2}
=
R_{\infty}^{2}\left(2du^{\prime} dv^{\prime\prime}\rho^{2}
-3 q_{1} du^{\prime\, 2}-\frac{d\rho^{2}}{\rho^{2}}\right)
-R_{\infty}^{2}d\Omega_{(3)}^{2}\, ,
\end{equation}
 
\noindent
where $u= R_{\infty}^{2} u^{\prime}$ and $v^{\prime}= R_{\infty}^{2}
v^{\prime\prime}$.

The transverse part of the metric is that of a round 3-sphere of radius
$R_{\infty}$. The rest turns out to be the metric of an AdS$_{3}$ space of radius
$R_{\infty}$ as well: computing its Riemann tensor we find

\begin{equation}
R^{(3)}_{\mu\nu\rho\sigma}
=
-\frac{2}{R_{\infty}^{2}}\, g^{(3)}_{\mu[\rho}g^{(3)}_{\sigma]\nu}\, .
\end{equation}

Thus, the second choice of integration constants gives a solutions which is
asymptotically AdS$_{3}\times S^{3}$ with radii equal to $R_{\infty}$. Observe
that, in the Abelian case (which we can always recover by eliminating the
instanton field) the solution would be globally, and not just asymptotically,
AdS$_{3}\times S^{3}$. In the $\rho\rightarrow \infty$ limit we recover
essentially this Abelian solution because the instanton field vanishes and, in
particular, the 3-form field strength $\breve{H}$ takes the form

\begin{equation}
\label{eq:limitbreveH}
\breve{H}
=
\tfrac{3}{2}q_{1}\left[ -\pi_{3}+\omega_{3}\right]\, ,
\end{equation}

\noindent
where $\pi_{3}$ and $\omega_{3}$ are the volume forms of unit-radii AdS$_{3}$
and $S^{3}$, respectively. In the coordinates we are using, the first is given by

\begin{equation}
\pi_{3} = \rho d\rho\wedge dv^{\prime\prime} \wedge du^{\prime}\, .  
\end{equation}

\end{enumerate}

Now we are interested in studying the near-horizon ($\rho\rightarrow 0$)
limits of these two solutions.

\subsubsection*{Near-horizon limit}

For any values of the integration constants $a_{0},a_{1}$ (that is: for the
two different solutions identified above), in the limit $\rho\rightarrow 0$,
the Ricci scalar and the Kretschmann invariant of the full metric remain
finite Thus, we expect to have a well-defined $\rho\rightarrow 0$ metric which
in the asymptotically-flat case will be interpreted as a near-horizon metric.
In both cases we have the the following asymptotic expansions:

\begin{equation}
L_{0,1} \sim \frac{q_{0,1}}{\rho^{2}}+\mathcal{O}(1), \quad
\breve{f}=\rho^{2}/R_{\rm h}^{2}+\mathcal{O}(\rho^4)\, ,
\end{equation}

\noindent
where\footnote{Compare this expression with Eq.~(\ref{eq:Rinfty}).} 

\begin{equation}
\label{eq:Rh}
R_{\rm h}^{2} 
\equiv
\sqrt{\frac{9q_{1}(q_{0}-8/(3\breve{g}^{2}))}{2}}\, ,
\end{equation}

\noindent
which is well defined as long as $q_{1}(q_{0}-8/(3\breve{g}^{2}))>0$ (in
particular, $q_{1}\neq 0$). We will assume that this condition holds. Then,
rescaling the null coordinates as $u=R_{\rm h}^{2}u^{\prime}$, $v^{\prime}=
R_{\rm h}^{2} v^{\prime\prime}$ the metric takes the same form we found above

\begin{equation}
d\breve{s}_{\rm h}^{2}
=
R_{\rm h}^{2}\left(2\rho^{2} du^{\prime} dv^{\prime\prime}
-3 q_{1} du^{\prime\, 2}-\frac{d\rho^{2}}{\rho^{2}}\right)
-R_{\rm h}^{2}d\Omega_{(3)}^{2}\, ,
\end{equation}

\noindent
which is that of AdS$_{3}\times S^{3}$ with radii equal to $R_{\rm h}$. The
fact that this near-horizon limit is the same as in the acase of the pure
dyonic string solutions (with no $pp$-wave) \cite{Duff:1998cr} is somewhat
surprising.

In this limit the dilaton takes a constant and finite value,

\begin{equation}
e^{\sqrt{2} \breve{\varphi}} 
= 
\frac{q_{1}}{2(q_{0}-\frac{8}{3\breve{g}^{2}})}\, ,
\end{equation}

\noindent
while the vectors are simply proportional to the left-invariant Maurer-Cartan
1-forms $\breve{A}^{A} = -\frac{1}{\breve{g}} \, v^{A}{}_{L}$. Recalling the
definition of the left-invariant Maurer-Cartan forms $V=v^{A}T_{A}= -u^{-1}du$
for the $\mathrm{SU}(2)$ group representative $u$ and the $\mathfrak{su}(2)$
generators $T_{A}$, we conclude that the gauge fields are proportional to a
pure gauge configuration, i.e.~they describe a meron field, analogous to the
one found in Ref.~\cite{Cariglia:2004kk}. Finally, in the $\rho\rightarrow 0$
limit the 3-form field strength $\breve{H}$ takes exactly the same form as in
the $\rho\rightarrow \infty$ limit Eq.~(\ref{eq:limitbreveH}), but we should
notice that the coordinates we are using in the AdS$_{3}$ are different.

Summarizing, we have found two solutions:

\begin{enumerate}
\item The first solution, which is asymptotically flat and has a regular
  horizon. Asymptotically it cannot be distinguished from the well-known
  dyonic string solution (plus $pp$-wave) that one can obtain by eliminating
  the instanton field. This behaviour is similar to that of the colored black
  holes constructed in
  Refs.~\cite{Meessen:2008kb,Meessen:2015nla,Meessen:2015enl}. In the
  near-horizon limit it has an AdS$_{3}\times S^{3}$ metric with radius
  $R_{\rm h}$ whose value, given in Eq.~(\ref{eq:Rh}) does have a contribution
  from the instanton field.

\item The second solution is a globally regular metric that interpolates
  between two AdS$_{3}\times S^{3}$ solutions with radii $R_{\infty}$ and
  $R_{h}$ given, respectively, in Eq.~(\ref{eq:Rinfty}) and Eq.~(\ref{eq:Rh}).

\end{enumerate}

We will discuss these solutions further in the Conclusions Section.

\subsection{Solutions of the  $\mathrm{SO}(3)$-gauged $\mathcal{N}=2A,d=6$ theory}

Dualizing the 3-form field strength of the $\mathcal{N}=2A^{*},d=6$ theory
solutions we just obtained according to Eq.~(\ref{eq:3formduality}) we can get
very similar solutions of the $\mathcal{N}=2A,d=6$ theory which will have,
however, very different string-frame metrics and (possibly) Kalb-Ramond field.

\begin{equation}
\tilde{H}
=
-\tfrac{1}{3} dv\wedge du \wedge d L_{1}^{-1} 
-\tfrac{3}{2} \rho^{3}\partial_{\rho} \left[L_{0}- \tfrac{2}{9g^{2}} \, 
\left(\rho+\frac{\lambda^{2}}{4} \rho^{3}\right)^{-2}\right] \omega_{3}\, .
\end{equation}

Since, in this case, the 3- and 2-form field strengths are defined as

\begin{eqnarray}
\tilde{H} 
& = &
d\tilde{B}  +\tilde{F}^{A}\wedge \tilde{A}^{A} 
+\tfrac{1}{3!}\tilde{g}\varepsilon_{ABC}
\tilde{A}^{A}\wedge\tilde{A}^{B}\wedge\tilde{A}^{C}\, ,
\\
& & \nonumber \\
\tilde{F}^{A}
& = &
d \tilde{A}^{A}
-\tfrac{1}{2}\tilde{g}\varepsilon^{A}{}_{BC}
\tilde{A}^{B}\wedge\tilde{A}^{C}\, ,
\end{eqnarray}

\noindent
and the gauge fields are those of the BPS instanton

\begin{equation}
\tilde{A}^{A} =- \frac{1}{\tilde{g}}\frac{1}{1+\tfrac{\lambda^{2}}{4}\rho^{2}}
v^{A}{}_{L}\, ,  
\end{equation}

\noindent
we find that

\begin{equation}
d\tilde{B}
=
-\tfrac{1}{3} dv\wedge du \wedge d L_{1}^{-1} 
+
3q_{0}\omega_{3}\, ,  
\end{equation}

\noindent
and using the Euler coordinates as in Eq.~(\ref{eq:BreveB}), we obtain the
2-form field

\begin{equation}
\tilde{B} 
=
-\tfrac{1}{3}L_{1}^{-1} dv\wedge du  
+
\tfrac{3}{8}q_{0}\cos{\theta} d\psi\wedge d\phi\, ,
\end{equation}

\noindent
which has no non-Abelian contributions.

\subsection{Solutions of the  ``$\mathrm{SO}(3)$-gauged'' $\mathcal{N}=2B,d=6$ theory}

As we have already mentioned, there is no possible gauging in any conventional
sense of the $\mathcal{N}=2B,d=6$ supergravity theory because it has no vector
fields. However, it can be argued that, at least when the theory is
compactified in a circle, a gauged $\mathcal{N}=2B,d=6$ supergravity theory
exists whose massless (in the 5-dimensional sense) sector is given by a gauged
$\mathcal{N}=2,d=5$ theory related to the former by dimensional reduction in
the Abelian case.

We have also stressed that the relation between the fields of two gauged
supergravities is the same as in the ungauged case, as long as their gauge
groups are identical. Then, we can use the formulae obtained in the
dimensional reduction of the standard $\mathcal{N}=2B,d=6$ to ungauged
$\mathcal{N}=2,d=5$ supergravity to uplift solutions of the
$\mathrm{SO}(3)$-gauged 5-dimensional theory to this conjectured
$\mathrm{SO}(3)$- gauged $\mathcal{N}=2B,d=6$ supergravity. We are going to
apply this idea to the non-Abelian black-hole solution we have uplifted to the
gauged $\mathcal{N}=2A$ and $\mathcal{N}=2A^{*},d=6$ theories. Eliminating the
BPST instanton from the solution we obtain a solution of the standard
(ungauged) $\mathcal{N}=2B,d=6$ theory.

Thus, using Eqs.~(\ref{scalarhat}),(\ref{metrichat}),(\ref{2formhat}), calling
$u$ and $v$ the coordinates $z$ and $t$ and shifting $v^{\prime}=v+3a_{0}u$ we
get the following solution 

\begin{equation}
\begin{array}{rcl}
d\hat{s}^{2}  
& = &  
\left({\displaystyle\frac{2}{3 L_{1}}}\right) 2du 
\left\{ 
dv^{\prime} -3 \left[(L_{0}-a_{0})- \tfrac{2}{9g^{2}} \, 
\left(\rho+\frac{\lambda^{2}}{4} \rho^{3}\right)^{-2}\right]du 
\right\}
\\
& & \\
& &  
- \left({\displaystyle\frac{2}{3 L_{1}}}\right)^{-1} 
\left( d \rho^{2} + \rho^{2} d \Omega_{(3)}^{2} \right)  \, ,
\\
& & \\ 
\hat{L}^{r}  
& = &  
\delta^{r}{}_{1}\, , 
\\
& & \\
\hat{B}^{1}{}_{uv^{\prime}}  
& = &  
\tfrac{1}{3} L_{1}^{-1} \, , 
\\
& & \\
\hat{B}^{A}{}_{ \mu u}dx^{\mu}  
& = &  
-\frac{1}{2\sqrt{6}g}\, v^{A}{}_{L} \, .
\end{array}
\end{equation}

This solution has the typical form of a solution describing the superposition
of a self-dual string with charge $\sim q_{1}$ and a $pp$-wave with momentum
$\sim q_{0}$ but there is a non-conventional non-Abelian contribution to this
wave which can be interpreted as an instanton expressed in 2-form
variables. This non-Abelian contribution, as in the previous cases, falls off
too fast at infinity to give a contribution to the wave's momentum and,
therefore, the solution has the same asymptotic behaviour as the standard
solution with no non-Abelian contribution. It also seems to be regular
everywhere as long as $L_{1}\neq 0$ (but we always choose $a_{1}$ and $q_{1}$
with equal signs).

In this solution the string charge and the $pp$-wave momentum are independent
and can be set to zero independently.Setting both to zero gives a
non-standard, purely non-Abelian $pp$-wave solution.

\subsubsection*{Asymptotic limit}

There are two possible choices of the integration constant $a_{1}$ which give
physically inequivalent solutions:\footnote{Observe that $a_{0}$ has
  disappeared from the solution.}

\begin{enumerate}
\item $a_{1}=1$ gives an asymptotically ($\rho\rightarrow \infty$ limit) flat
    metric with the string-plus-wave interpretation mentioned above.
  \item $a_{1}=0$ gives a metric that, with the usual rescaling of $u$ and
    $v^{\prime}$, takes the form

\begin{equation}
d\hat{s}^{2}
=
R^{2}
\left\{
\left[
2du^{\prime}dv^{\prime\prime}\rho^{2} 
-3\left(q_{0} -\tfrac{2}{9g^{2}}(1+\tfrac{\lambda^{2}}{4}\rho^{2})^{-2}\right)du^{\prime}
-\frac{d\rho^{2}}{\rho^{2}}
\right]
-d\Omega^{2}_{(3)}
\right\}\, .      
\end{equation}

In the $\rho\rightarrow \infty$ limit this metric is that of AdS$_{3}\times$S$^{3}$
with radii

\begin{equation}
\label{eq:R}
R^{2}= 3q_{1}/2\, ,   
\end{equation}

\noindent
but, for all finite values of $\rho$ it is different from it, except when the
non-Abelian contribution is eliminated.

\end{enumerate}

\subsubsection*{Near-horizon limit}

For the two solutions $a_{1}=1,0$ one obtains the same metric in the
$\rho\rightarrow 0$ (near-horizon) limit: an AdS$_{3}\times$S$^{3}$
with radii $R$ given by Eq.~(\ref{eq:R}). The difference between this metric
and the one obtained in the $\rho\rightarrow \infty$ limit for the second
solution is that in the near-horizon limit there is a non-Abelian
contribution in the $g_{uu}$ component, although this does not affect the value
of the radii of the factor spaces.

\section{Conclusions}
\label{sec-conclusions}

We have found a very interesting relation between two families of models of
$\mathcal{N}=(2,0),d=6$ supergravity that can be used to transform solutions
of one of them admitting one isometry into solutions of the other. The
relation is based on the fact that they reduce to the same family of models of
$\mathcal{N}=2,d=5$ supergravity, a fact that we have used to construct new
6-dimensional supersymmetric non-Abelian solutions by uplifting a known
5-dimensional solution. 

It is natural to expect that the relation between 6-dimensional supergravities
is related to a string duality, but more work is necessary in order to
identify the string compactifications that produce the 6-dimensional theories
that only have chiral 2-forms. 

We have only uplifted the simplest non-Abelian 5-dimensional solution (a black
hole), but one should consider more possibilities like the non-Abelian black
ring or rotating black hole of Ref.~\cite{Ortin:2016bnl}. As in the 5- and
4-dimensional cases, the non-Abelian does not contribute to any of the
quantities one can measure at infinity, like the mass, but it does modify the
near-horizon geometry, with a negative contribution to the entropy. This means
that, for the same asymptotic data there are several black-body configurations
with different entropies and the non-Abelian one, having the least entropy,
should be unstable. An intriguing possibility is that the solution that
interpolates between two different AdS$_{3}\times$S$_{3}$ geometries is
somehow related to an instanton associated to that instability. Work in this
direction is underway \cite{kn:COS}.

Finally, a long-standing problem that remains unsolved as yet is the
microscopical interpretation of the entropy of all the black objects with
non-Abelian fields found so far. We believe that the work presented here will
help to find the embedding of these solutions in a string theory, providing
the first step to solve it.

\section*{Acknowledgments}

T.O.~would like to thank Patrick Meessen for many useful conversations.  This
work has been supported in part by the Spanish Ministry of Science and
Education grants FPA2012-35043-C02-01 and FPA2015-66793-P, the Centro de
Excelencia Severo Ochoa Program grant SEV-2012-0249.  The work of P.A.C.~was
supported by a ``la Caixa-Severo Ochoa'' International pre-doctoral grant. TO
wishes to thank M.M.~Fern\'andez for her permanent support.

\appendix



\begin{thebibliography}{99}

\bibitem{Ferrara:1995ih}
S.~Ferrara, R.~Kallosh and A.~Strominger,
``$\mathcal{N}=2$ extremal black holes,''
Phys.\ Rev.\ D {\bf 52} (1995) 5412.
\doi{10.1103/PhysRevD.52.R5412}.
[\hepth{9508072}].

\bibitem{Strominger:1996kf}
A.~Strominger,
``Macroscopic Entropy of $\mathcal{N}=2$ Extremal Black Holes,''
Phys.\ Lett.\ B {\bf 383} (1996) 39.
\doi{10.1016/0370-2693(96)00711-3}.
[\hepth{9602111}].

\bibitem{Ferrara:1996dd}
S.~Ferrara and R.~Kallosh,
``Supersymmetry and Attractors,''
Phys.\ Rev.\ D {\bf 54} (1996) 1514.
\doi{10.1103/PhysRevD.54.1514}.
[\hepth{9602136}].

\bibitem{Ferrara:1996um}
S.~Ferrara and R.~Kallosh,
``Universality of Supersymmetric Attractors,''
Phys.\ Rev.\ D {\bf 54} (1996) 1525.
\doi{10.1103/PhysRevD.54.1525}.
[\hepth{9603090}].

\bibitem{Ferrara:1997tw} 
S.~Ferrara, G.\,W.~Gibbons, R.~Kallosh,
``Black holes and critical points in moduli space,''
Nucl.\ Phys.\ B {\bf 500} (1997) 75.
\doi{10.1016/S0550-3213(97)00324-6}.
[\hepth{9702103}].

\bibitem{Freedman:2012zz}
D.~Z.~Freedman and A.~Van Proeyen,
``Supergravity,''
Cambridge, UK: Cambridge Univ. Pr. (2012) 607 p

\bibitem{Andrianopoli:1996cm}
L.~Andrianopoli, M.~Bertolini, A.~Ceresole, R.~D'Auria, S.~Ferrara,
P.~Fr\'e and T.~Magri,
``N=2 supergravity and N=2 superYang-Mills theory on general scalar manifolds: Symplectic covariance, gaugings and the momentum map,''
J.\ Geom.\ Phys.\  {\bf 23} (1997) 111.
\doi{10.1016/S0393-0440(97)00002-8}.
[\hepth{9605032}].

\bibitem{Ortin:2015hya}
T.~Ort\'{\i}n,
``Gravity and Strings'', 2nd edition, 
Cambridge University Press, 2015.

\bibitem{Tod:1983pm}
K.P.~Tod,
``All Metrics Admitting Supercovariantly Constant Spinors,''
Phys.\ Lett.\ B {\bf 121} (1983) 241.
\doi{10.1016/0370-2693(83)90797-9}.

\bibitem{Caldarelli:2003pb}
M.M.~Caldarelli and D.~Klemm,
 ``All supersymmetric solutions of N = 2, D = 4 gauged supergravity,''
JHEP {\bf 0309} (2003) 019.
\doi{10.1088/1126-6708/2003/09/019}.
[\hepth{0307022}].

\bibitem{Meessen:2006tu}
P.~Meessen and T.~Ort\'{\i}n,
``The Supersymmetric configurations of N=2, D=4 supergravity coupled to vector supermultiplets,''
Nucl.\ Phys.\ B {\bf 749} (2006) 291.
\doi{10.1016/j.nuclphysb.2006.05.025}.
[\hepth{0603099}].

\bibitem{Huebscher:2006mr}
M.~H\"ubscher, P.~Meessen and T.~Ort\'{\i}n,
``Supersymmetric solutions of N=2 D=4 sugra: The Whole ungauged shebang,''
Nucl.\ Phys.\ B {\bf 759} (2006) 228.
\doi{10.1016/j.nuclphysb.2006.10.004}.
[\hepth{0606281}].

\bibitem{Cacciatori:2008ek}
S.~L.~Cacciatori, D.~Klemm, D.~S.~Mansi and E.~Zorzan,
``All timelike supersymmetric solutions of N=2, D=4 gauged supergravity
coupled to abelian vector multiplets,''
JHEP {\bf 0805} (2008) 097.
\doi{10.1088/1126-6708/2008/05/097}.
[\arxiv{0804.0009} [hep-th]].

\bibitem{Hubscher:2008yz}
M.~H\"ubscher, P.~Meessen, T.~Ort\'{\i}n and S.~Vaul\`a,
``N=2 Einstein-Yang-Mills's BPS solutions,''
JHEP {\bf 0809} (2008) 099.
\doi{10.1088/1126-6708/2008/09/099}.
[\arxiv{0806.1477} [hep-th]].

\bibitem{Meessen:2012sr}
P.~Meessen and T.~Ort\'{\i}n,
``Supersymmetric solutions to gauged $N=2$ $d=4$ sugra: the full timelike shebang,''
Nucl.\ Phys.\ B {\bf 863} (2012) 65.
\doi{10.1016/j.nuclphysb.2012.05.023}.
[\arxiv{1204.0493}].

\bibitem{Gauntlett:2002nw}
J.P.~Gauntlett, J.B.~Gutowski, C.M.~Hull, S.~Pakis and H.S.~Reall,
``All supersymmetric solutions of minimal supergravity in five dimensions,''
Class.\ Quant.\ Grav.\  {\bf 20} (2003) 4587.
\doi{10.1088/0264-9381/20/21/005}.
[\hepth{0209114}].

\bibitem{Gauntlett:2003fk}
J.~P.~Gauntlett and J.~B.~Gutowski,
``All supersymmetric solutions of minimal gauged supergravity in five-dim
ensions,''
Phys.\ Rev.\ D {\bf 68} (2003) 105009.
Erratum: [Phys.\ Rev.\ D {\bf 70} (2004) 089901].
\doi{10.1103/PhysRevD.70.089901}, \doi{10.1103/PhysRevD.68.105009}.
[\hepth{0304064}].

\bibitem{Gauntlett:2004qy}
J.P.~Gauntlett and J.B.~Gutowski,
``General concentric black rings,''
Phys.\ Rev.\ D {\bf 71} (2005) 045002.
\doi{10.1103/PhysRevD.71.045002}.
[\hepth{0408122}].

\bibitem{Gutowski:2004yv}
J.~B.~Gutowski and H.~S.~Reall,
``General supersymmetric AdS(5) black holes,''
JHEP {\bf 0404} (2004) 048
\doi{10.1088/1126-6708/2004/04/048}
[\hepth{0401129}].

\bibitem{Gutowski:2005id}
J.~B.~Gutowski and W.~Sabra,
``General supersymmetric solutions of five-dimensional supergravity,''
JHEP {\bf 0510} (2005) 039
\doi{10.1088/1126-6708/2005/10/039}
[\hepth{0505185}].

\bibitem{Bellorin:2006yr}
J.~Bellor\'{\i}n, P.~Meessen and T.~Ort\'{\i}n,
``All the supersymmetric solutions of N=1,d=5 ungauged supergravity,''
JHEP {\bf 0701} (2007) 020.
\doi{10.1088/1126-6708/2007/01/020}.
[\hepth{0610196}].

\bibitem{Bellorin:2007yp}
J.~Bellor\'{\i}n and T.~Ort\'{\i}n,
``Characterization of all the supersymmetric solutions of gauged N=1, d=5 supergravity,''
JHEP {\bf 0708} (2007) 096
\doi{10.1088/1126-6708/2007/08/096}.
[\arxiv{0705.2567} [hep-th]].

\bibitem{Bellorin:2008we}
J.~Bellor\'{\i}n,
``Supersymmetric solutions of gauged five-dimensional supergravity with general matter couplings,''
Class.\ Quant.\ Grav.\  {\bf 26} (2009) 195012.
\doi{10.1088/0264-9381/26/19/195012}.
[\arxiv{0810.0527} [hep-th]].

\bibitem{Marcus:1982yu}
N.~Marcus and J.~H.~Schwarz,
``Field Theories That Have No Manifestly Lorentz Invariant Formulation,''
Phys.\ Lett.\ B {\bf 115} (1982) 111.
\doi{10.1016/0370-2693(82)90807-3}.

\bibitem{Cremmer:1978km}
E.~Cremmer, B.~Julia and J.~Scherk,
``Supergravity Theory in Eleven-Dimensions,''
Phys.\ Lett.\ B {\bf 76} (1978) 409.
\doi{10.1016/0370-2693(78)90894-8}

\bibitem{Nishino:1984gk}
H.~Nishino and E.~Sezgin,
``Matter and Gauge Couplings of N=2 Supergravity in Six-Dimensions,''
Phys.\ Lett.\ B {\bf 144} (1984) 187.
\doi{10.1016/0370-2693(84)91800-8}

\bibitem{Bergshoeff:1985mz}
E.~Bergshoeff, E.~Sezgin and A.~Van Proeyen,
``Superconformal Tensor Calculus and Matter Couplings in Six-dimensions,''
Nucl.\ Phys.\ B {\bf 264} (1986) 653.
Erratum: [Nucl.\ Phys.\ B {\bf 598} (2001) 667].
\doi{10.1016/0550-3213(86)90503-1}.

\bibitem{Nishino:1986dc}
H.~Nishino and E.~Sezgin,
``The Complete $N=2$, $d=6$ Supergravity With Matter and {Yang-Mills} Couplings,''
Nucl.\ Phys.\ B {\bf 278} (1986) 353.
\doi{10.1016/0550-3213(86)90218-X}

\bibitem{Romans:1986er}
L.~J.~Romans,
``Selfduality for Interacting Fields: Covariant Field Equations for Six-dimensional Chiral Supergravities,''
Nucl.\ Phys.\ B {\bf 276} (1986) 71.
\doi{10.1016/0550-3213(86)90016-7}

\bibitem{Nishino:1997ff}
H.~Nishino and E.~Sezgin,
``New couplings of six-dimensional supergravity,''
Nucl.\ Phys.\ B {\bf 505} (1997) 497.
\doi{10.1016/S0550-3213(97)00357-X}.
[\hepth{9703075}].

\bibitem{Gutowski:2003rg} 
J.~B.~Gutowski, D.~Martelli and H.S.~Reall,
``All supersymmetric solutions of minimal supergravity in six dimensions'',
Class.~Quant.~Grav.~\textbf{20} (2003) 5049.
\doi{10.1088/0264-9381/20/23/008}.
[\hepth{0306235}].

\bibitem{Chamseddine:2003yy} 
A.\,H.~Chamseddine, J.\,M.~Figueroa-O'Farrill and W.~Sabra,
``Six-Dimensional Supergravity Vacua and Anti-Selfdual Lorentzian Lie Groups,''
\hepth{0306278}.

\bibitem{Cariglia:2004kk}
M.~Cariglia and O.~A.~P.~Mac Conamhna,
``The General form of supersymmetric solutions of N=(1,0) U(1) and SU(2)
gauged supergravities in six-dimensions,''
Class.\ Quant.\ Grav.\  {\bf 21} (2004) 3171.
\doi{10.1088/0264-9381/21/13/006}.
[\hepth{0402055}].

\bibitem{kn:CMOT}
P.A.~Cano, P.~Meessen, T.~Ort\'{\i}n and E.~Torrente-Luj\'an,
work in preparation.

\bibitem{Bergshoeff:1994cb}
E.~Bergshoeff, R.~Kallosh, and T.~Ort\'{\i}n,
``Duality Versus Supersymmetry and Compactification,''
Phys.\ Rev.\ D {\bf 51} (1995) 3009.
\doi{10.1103/PhysRevD.51.3009}.
[\hepth{9410230}].

\bibitem{Bergshoeff:1995as}
E.~Bergshoeff, C.~M.~Hull and T.~Ort\'{\i}n,
``Duality in the type II superstring effective action,''
Nucl.\ Phys.\ B {\bf 451} (1995) 547
\doi{10.1016/0550-3213(95)00367-2}.
[\hepth{9504081}].

\bibitem{Dai:1989ua}
J.~Dai, R.~G.~Leigh and J.~Polchinski,
``New Connections Between String Theories,''
Mod.\ Phys.\ Lett.\ A {\bf 4} (1989) 2073.
\doi{10.1142/S0217732389002331}.

\bibitem{Dine:1989vu}
M.~Dine, P.~Y.~Huet and N.~Seiberg,
``Large and Small Radius in String Theory,''
Nucl.\ Phys.\ B {\bf 322} (1989) 301.
\doi{10.1016/0550-3213(89)90418-5}

\bibitem{Witten:1995ex}
E.~Witten,
``String theory dynamics in various dimensions,''
Nucl.\ Phys.\ B {\bf 443} (1995) 85.
\doi{10.1016/0550-3213(95)00158-O}.
[\hepth{9503124}].

\bibitem{Buscher:1985kb}
T.~H.~Buscher,
``Quantum Corrections and Extended Supersymmetry in New $\sigma$ Models,''
Phys.\ Lett.\ B {\bf 159} (1985) 127.
\doi{10.1016/0370-2693(85)90870-6}.

\bibitem{Buscher:1987sk}
T.~H.~Buscher,
``A Symmetry of the String Background Field Equations,''
Phys.\ Lett.\ B {\bf 194} (1987) 59.
\doi{10.1016/0370-2693(87)90769-6}.

\bibitem{Buscher:1987qj}
T.~H.~Buscher,
``Path Integral Derivation of Quantum Duality in Nonlinear Sigma Models,''
Phys.\ Lett.\ B {\bf 201} (1988) 466.
\doi{10.1016/0370-2693(88)90602-8}.

\bibitem{Hofman:2002ey}
C.~Hofman,
``NonAbelian 2 forms,''
\hepth{0207017}.

\bibitem{Ho:2011ni}
P.~M.~Ho, K.~W.~Huang and Y.~Matsuo,
``A Non-Abelian Self-Dual Gauge Theory in 5+1 Dimensions,''
JHEP {\bf 1107} (2011) 021.
\doi{10.1007/JHEP07(2011)021}
[\arxiv{1104.4040} [hep-th]].

\bibitem{Huang:2012tu}
K.~W.~Huang,
``Non-Abelian Chiral 2-Form and M5-Branes,''
\arxiv{1206.3983} [hep-th].

\bibitem{Meessen:2015enl}
P.~Meessen, T.~Ort\'{\i}n and P.~Fern\'andez-Ram\'{\i}rez,
``Non-Abelian, supersymmetric black holes and strings in 5 dimensions,''
JHEP {\bf 1603} (2016) 112.
\doi{10.1007/JHEP03(2016)112}.
[\arxiv{1512.07131} [hep-th]].

\bibitem{Ortin:2016bnl}
T.~Ort\'{\i}n and P.~F.~Ram\'{\i}rez,
``A non-Abelian Black Ring,''
\arxiv{1605.00005} [hep-th].
To be publishedin JHEP

\bibitem{Huebscher:2007hj}
M.~H\"ubscher, P.~Meessen, T.~Ort\'{\i}n and S.~Vaul\`a,
``Supersymmetric N=2 Einstein-Yang-Mills monopoles and covariant attractors,''
Phys.\ Rev.\ D {\bf 78} (2008) 065031.
\doi{10.1103/PhysRevD.78.065031}.
[\arxiv{0712.1530} [hep-th]].

\bibitem{Meessen:2008kb}
P.~Meessen,
``Supersymmetric coloured/hairy black holes,''
Phys.\ Lett.\ B {\bf 665} (2008) 388.
\doi{10.1016/j.physletb.2008.06.035}.
[\arxiv{0803.0684} [hep-th]].

\bibitem{Meessen:2015nla}
P.~Meessen and T.~Ort\'{\i}n,
``$\mathcal{N}=2 $ super-EYM coloured black holes from defective Lax matrices,''
JHEP {\bf 1504} (2015) 100.
\doi{10.1007/JHEP04(2015)100}.
[\arxiv{1501.02078} [hep-th]].

\bibitem{Bueno:2014mea}
P.~Bueno, P.~Meessen, T.~Ort\'{\i}n and P.~F.~Ram\'{\i}rez,
``$ \mathcal{N}=2 $ Einstein-Yang-Mills' static two-center solutions,''
JHEP {\bf 1412} (2014) 093.
\doi{10.1007/JHEP12(2014)093}.
[\arxiv{1410.4160} [hep-th]].

\bibitem{Bueno:2015wva}
P.~Bueno, P.~Meessen, T.~Ort\'{\i}n and P.~F.~Ram\'{\i}rez,
``Resolution of SU(2) monopole singularities by oxidation,''
Phys.\ Lett.\ B {\bf 746} (2015) 109.
\doi{10.1016/j.physletb.2015.04.065}.
[\arxiv{1503.01044} [hep-th]].

\bibitem{Bergshoeff:1995sq}
E.~Bergshoeff, H.~J.~Boonstra and T.~Ort\'{\i}n,
``S duality and dyonic p-brane solutions in type II string theory,''
Phys.\ Rev.\ D {\bf 53} (1996) 7206.
\doi{10.1103/PhysRevD.53.7206}
[\hepth{9508091}].

\bibitem{Schwarz:1983wa}
J.~H.~Schwarz and P.~C.~West,
``Symmetries and Transformations of Chiral N=2 D=10 Supergravity,''
Phys.\ Lett.\ B {\bf 126} (1983) 301.
\doi{10.1016/0370-2693(83)90168-5}.

\bibitem{Schwarz:1983qr}
J.~H.~Schwarz,
``Covariant Field Equations of Chiral N=2 D=10 Supergravity,''
Nucl.\ Phys.\ B {\bf 226} (1983) 269.
\doi{10.1016/0550-3213(83)90192-X}.

\bibitem{Howe:1983sra}
P.~S.~Howe and P.~C.~West,
``The Complete N=2, D=10 Supergravity,''
Nucl.\ Phys.\ B {\bf 238} (1984) 181.
\doi{10.1016/0550-3213(84)90472-3}.

\bibitem{Meessen:1998qm}
P.~Meessen and T.~Ort\'{\i}n,
``An Sl(2,Z) multiplet of nine-dimensional type II supergravity theories,''
Nucl.\ Phys.\ B {\bf 541} (1999) 195.
\doi{10.1016/S0550-3213(98)00780-9}.
[\hepth{9806120}].

\bibitem{Hartong:2009vc}
J.~Hartong and T.~Ort\'{\i}n,
``Tensor Hierarchies of 5- and 6-Dimensional Field Theories,''
JHEP {\bf 0909} (2009) 039.
\doi{10.1088/1126-6708/2009/09/039}.
[\arxiv{0906.4043} [hep-th]].

\bibitem{Scherk:1979zr}
J.~Scherk and J.~H.~Schwarz,
``How to Get Masses from Extra Dimensions,''
Nucl.\ Phys.\ B {\bf 153} (1979) 61.
\doi{10.1016/0550-3213(79)90592-3}.

\bibitem{Bergshoeff:2004kh}
E.~Bergshoeff, S.~Cucu, T.~de Wit, J.~Gheerardyn, S.~Vandoren and A.~Van Proeyen,
``N = 2 supergravity in five-dimensions revisited,''
Class.\ Quant.\ Grav.\  {\bf 21} (2004) 3015.
[Class.\ Quant.\ Grav.\  {\bf 23} (2006) 7149].
\doi{10.1088/0264-9381/23/23/C01}, \doi{10.1088/0264-9381/21/12/013}.
[\hepth{0403045}].

\bibitem{Meessen:2001wk}
P.~Meessen and T.~Ort\'{\i}n,
``Type 0 T duality and the tachyon coupling,''
Phys.\ Rev.\ D {\bf 64} (2001) 126005
\doi{10.1103/PhysRevD.64.126005}.
[\hepth{0103244}].

\bibitem{Seiberg:1988pf}
N.~Seiberg,
``Observations on the Moduli Space of Superconformal Field Theories,''
Nucl.\ Phys.\ B {\bf 303} (1988) 286.
\doi{10.1016/0550-3213(88)90183-6}.

\bibitem{Duff:1993ij}
M.~J.~Duff and R.~R.~Khuri,
``Four-dimensional string / string duality,''
Nucl.\ Phys.\ B {\bf 411} (1994) 473.
\doi{10.1016/0550-3213(94)90459-6}.
[\hepth{9305142}].

\bibitem{Hull:1994ys}
C.~M.~Hull and P.~K.~Townsend,
``Unity of superstring dualities,''
Nucl.\ Phys.\ B {\bf 438} (1995) 109.
\doi{10.1016/0550-3213(94)00559-W}
[\hepth{9410167}].

\bibitem{Duff:1994zt}
M.~J.~Duff,
``Strong / weak coupling duality from the dual string,''
Nucl.\ Phys.\ B {\bf 442} (1995) 47.
\doi{10.1016/S0550-3213(95)00070-4}.
[\hepth{9501030}].

\bibitem{Belavin:1975fg}
A.~A.~Belavin, A.~M.~Polyakov, A.~S.~Schwartz and Y.~S.~Tyupkin,
``Pseudoparticle Solutions of the Yang-Mills Equations,''
Phys.\ Lett.\ B {\bf 59} (1975) 85.
\doi{10.1016/0370-2693(75)90163-X}.

\bibitem{Duff:1996cf}
M.~J.~Duff, H.~L\"u and C.~N.~Pope,
``Heterotic phase transitions and singularities of the gauge dyonic string,''
Phys.\ Lett.\ B {\bf 378} (1996) 101.
\doi{10.1016/0370-2693(96)00420-0}.
[\hepth{9603037}].

\bibitem{Duff:1995yh}
M.~J.~Duff, S.~Ferrara, R.~R.~Khuri and J.~Rahmfeld,
``Supersymmetry and dual string solitons,''
Phys.\ Lett.\ B {\bf 356} (1995) 479.
\doi{10.1016/0370-2693(95)00838-C}.
[\hepth{9506057}].

\bibitem{Duff:1997me}
M.~J.~Duff, J.~T.~Liu, H.~L\"u and C.~N.~Pope,
``Gauge dyonic strings and their global limit,''
Nucl.\ Phys.\ B {\bf 529} (1998) 137.
\doi{10.1016/S0550-3213(98)00367-8}.
[\hepth{9711089}].

\bibitem{Duff:1998cr}
M.~J.~Duff, H.~L\"u and C.~N.~Pope,
``AdS$_{3}\times$S$^{3}$ (un)twisted and squashed, and an O(2,2,$\mathbb{Z}$) 
multiplet of dyonic strings,''
Nucl.\ Phys.\ B {\bf 544} (1999) 145.
\doi{10.1016/S0550-3213(98)00810-4}.
[\hepth{9807173}].

\bibitem{kn:COS} 
P.A.~Cano, C.~Santoli and T.~Ort\'{\i}n, 
work in preparation.





\end{thebibliography}
\end{document}